\documentclass[11pt,a4paper]{article}
\pdfoutput=1
\usepackage{jheppub}
\usepackage{color}
\usepackage{graphicx}
\usepackage{wrapfig}
\usepackage{verbatim}
\usepackage{amsmath}
\usepackage{amssymb}
\usepackage{url}
\usepackage{bbold}
\usepackage{xspace}
\usepackage{slashed}
\usepackage{multirow}
\usepackage{threeparttable}
\usepackage{paralist}
\usepackage{color}
\usepackage{subcaption}
\usepackage{floatrow}
\newfloatcommand{capbtabbox}{table}[][\FBwidth]

\usepackage{tikz}
\usetikzlibrary{trees}
\usetikzlibrary{decorations.pathmorphing}
\usetikzlibrary{decorations.markings}
\usetikzlibrary{arrows}

\definecolor{darkgreen}{rgb}{0,0.5,0}

\newcommand{\GeV}{\text{ GeV}}

\newcommand{\MeV}{\text{ MeV}}

\newcommand{\vev}[1]{\langle #1 \rangle}
\newcommand{\hc}{\text{h.c.}}

\newcommand{\PQ}{\text{\tiny PQ}}
\newcommand{\muI}{\mu}%\mu_I

\newcommand{\IR}{\text{\tiny IR}}
\newcommand{\KSVZ}{\text{\tiny KSVZ}}
\newcommand{\CFL}{\text{\tiny CFL}}
\newcommand{\EM}{\text{\tiny EM}}

\DeclareRobustCommand{\Sec}[1]{Sec.~\ref{#1}}

\DeclareRobustCommand{\App}[1]{App.~\ref{#1}}

\DeclareRobustCommand{\Fig}[1]{Fig.~\ref{#1}}

\DeclareRobustCommand{\Eq}[1]{Eq.~(\ref{#1})}
\DeclareRobustCommand{\Eqs}[2]{Eqs.~(\ref{#1}) and (\ref{#2})}
\DeclareRobustCommand{\EqsRange}[2]{Eqs.~(\ref{#1}) - (\ref{#2})}

\newcommand{\Dfbd}{\mathord{\buildrel{\lower3pt\hbox{$\scriptscriptstyle\leftrightarrow$}}\over {D}_{\mu}}}

\newcommand{\beq}{\begin{equation}}
\newcommand{\eeq}[1]{\label{#1}\end{equation}}
\def\beqa{\begin{eqnarray}}
\def\eeqa#1{\label{#1}\end{eqnarray}}
\newcommand{\eeqn}{\end{equation}}

\def\nn{\nonumber}

\def\Tr{\mathop{\rm Tr}}

\def\stacksymbols #1#2#3#4{\def\theguybelow{#2}
    \def\vp{\lower#3pt}
    \def\sp{\baselineskip0pt\lineskip#4pt}
    \mathrel{\mathpalette\intermediary#1}}

\def\intermediary#1#2{\vp\vbox{\sp
     \everycr={}\tabskip0pt
     \halign{$\mathsurround0pt#1\hfil##\hfil$\crcr#2\crcr
              \theguybelow\crcr}}}

\begin{document}
{\large
\flushright TUM-HEP-1255/20 \\
}
\title{The QCD Axion at Finite Density}
%\title{The Dense QCD Axion} 

\author{Reuven Balkin,}
\author{Javi Serra,}
\author{Konstantin Springmann,}
\author{Andreas Weiler}

\affiliation{Physik-Department, Technische Universit\"at M\"unchen, 85748 Garching, Germany}
\emailAdd{reuven.balkin@tum.de}
\emailAdd{javi.serra@tum.de}
\emailAdd{konstantin.springmann@tum.de}
\emailAdd{andreas.weiler@tum.de}

\date{\today}

\abstract{We show how the properties of the QCD axion change in systems at finite baryonic density, such as neutron stars. At nuclear saturation densities, where corrections can be reliably computed, we find a mild reduction of the axion mass and up to an order of magnitude enhancement in the model-independent axion coupling to neutrons. At moderately higher densities, if realized, meson (kaon) condensation can trigger axion condensation. We also study the axion potential at asymptotically large densities, where the color-superconducting phase of QCD potentially leads to axion condensation, and the mass of the axion is generically several orders of magnitude smaller than in vacuum due to the suppressed instantons. Several phenomenological consequences of the axion being sourced by neutron stars are discussed, such as its contribution to their total mass, the presence of an axionic brane, or axion-photon conversion in the magnetosphere.}

\preprint{}
\maketitle
\section{Introduction}

The QCD axion is one of the best motivated particles for physics beyond the Standard Model (SM). The Nambu-Goldstone boson~\cite{Weinberg:1977ma,Wilczek:1977pj} of a spontaneously broken $U(1)_{\PQ}$ Peccei-Quinn symmetry anomalous under QCD~\cite{Peccei:1977hh}, the axion makes the effective QCD $\theta$ angle unphysical and it allows QCD dynamics~\cite{Vafa:1984xg} to solve its strong-CP problem, thus explaining the absence of CP violation in the interactions of hadrons, in particular the electric dipole moment of the neutron~\cite{Crewther:1979pi,Abel:2020gbr}. In addition, the axion constitutes a viable and attractive  candidate for dark matter~\cite{Preskill:1982cy,Abbott:1982af,Dine:1982ah}.

The phenomenology of the axion is mainly controlled by the axion decay constant $f_a$. The defining coupling of the axion to the QCD topological charge determines its mass $m_a \approx m_\pi f_\pi/f_a \approx 5 \, \mu{\rm eV} (10^{12} \GeV/f_a)$~\footnote{This prediction seems however to be dependent on the UV dynamics of QCD, in particular on the size of the so-called small instanton contributions, see e.g.~\cite{Agrawal:2017ksf,Csaki:2019vte,Gherghetta:2020keg}. In this work we assume these are absent.} as well as its model-independent couplings to SM fields. Depending on the details of the axion UV completion, extra model-dependent derivative couplings can be present, which scale as well as $1/f_a$. These properties, and its dependence on $f_a$, endow the axion with a very interesting and diverse phenomenology, which is presently being actively investigated and for which there is a vigorous experimental effort, both current and planned, see e.g.~\cite{Irastorza:2018dyq}. Particularly relevant probes of the axion are astrophysical in nature, leading to a bound $f_a \gtrsim 10^8 \, \GeV$ from the cooling rate of the SN1987A supernova (see e.g.~\cite{Raffelt:2006cw} and references therein, as well as \cite{Chang:2018rso} and \cite{Bar:2019ifz} which have recently reevaluated and even put into question the feasibility of such a bound) and $f_a \lesssim 10^{17} \, \GeV$ from black hole superradiance~\cite{Arvanitaki:2009fg,Arvanitaki:2010sy,Arvanitaki:2014wva}.

In this paper we contribute to the global effort in the search for the QCD axion by presenting a first comprehensive study of how its properties change with chemical potential (and negligible temperature), of significant relevance for astrophysical systems, in particular those with large baryonic densities such as (proto-)neutron stars.

Indeed, neutron stars (NSs) are the densest stars in the universe, with densities that in their core are expected to go well beyond nuclear-saturation density, $n_0 \approx 0.16 \, {\rm nucleons}/{\rm fm}^3$ (i.e.~$\rho_0 \approx m_n n_0 \approx (190 \MeV)^4 \approx 3 \times 10^{14} \, {\rm g}/{\rm cm}^3$), which is also a density encountered in supernovae~\cite{Lattimer:2004pg,Lattimer:2012nd,Baym:2018aa}.
Given that such densities correspond to Fermi momenta not far from the scale of (strong) QCD dynamics, and in particular above the pion mass, it is clear that finite density corrections can have an important impact on the properties of the axion. As we will show, at such densities, where the QCD chiral Lagrangian at finite chemical potential is still reliable, the properties of the axion receive corrections of order $n/\Lambda_\chi f_\pi^2 \approx 20 \% \, (n/n_0)$, in particular leading to a reduction of the axion mass (taking $\Lambda_{\chi} \approx 700 \MeV$ to be the chiral Lagrangian cutoff). A notable exception is the model-independent coupling of the axion to neutrons, or just the neutron coupling in the KSVZ model, a.k.a.~hadronic axion~\cite{Kim:1979if,Shifman:1979if}. We find that the accidental cancellation present in vacuum (see e.g.~\cite{diCortona:2015ldu}) is spoiled at finite density, potentially resulting in an $O(10)$ enhancement, up to uncertainties of the relevant operators' coefficients and on yet to be computed extra corrections.

Since chiral perturbation theory (ChPT) is expected to fail beyond nuclear-saturation densities, and there is no other known reliable method to describe QCD at such high densities as those found in the core of NSs, we are driven to make extrapolations and qualitative statements about the possible fate of the QCD axion. However, under well defined assumptions, in particular that kaon condensation~\cite{Kaplan:1987sc} takes place somewhat above $n_0$, we show that spontaneous CP violation in the neutral Goldstone sector is triggered, resulting in axion condensation -- see~\cite{Hook:2017psm} for a first discussion of axion condensation in the context of a tuned non-standard axion. For a range of axion masses of order of the inverse radius of the NS, the axion gets sourced, with phenomenological consequences that go from a non-standard contribution to the mass of the star (closely resembling~\cite{Bellazzini:2015wva,Csaki:2018fls}) or the presence of an axionic brane~\cite{vacNS}, to an $O(1)$ enhancement of the electric field in the star's magnetosphere. Other implications of such a sourcing, e.g.~axion-mediated long-range forces with a potentially dramatic impact on NS mergers~\cite{Hook:2017psm}, do not seem to be realistic for the QCD axion.

To improve our understanding, at least at the qualitative level, of the physics of the axion at high densities, we also resort to the limit of very high quark chemical potential, $\mu_q \gg \Lambda_\chi$. 
In this regime, QCD is expected to reside in a color-flavor-locked (CFL) phase, where gluons are massive and a perturbative expansion is possible~\cite{Alford:1997zt,Rapp:1997zu,Alford:1998mk,Son:1998uk}. Our analysis of the axion potential in this regime shows that axion condensation is a generic phenomenon (as indirectly anticipated in~\cite{Kryjevski:2004cw}, where a non-zero effective $\theta$ angle was found), while the axion mass is orders of magnitude smaller than in vacuum, mainly due to the fact that QCD-instantons are very suppressed, scaling as $1/\mu_q^8$. The fact that we find the same qualitative behavior for both the edges of QCD perturbative control, i.e.~at mild and very high densities, is very suggestive of the fact that a non-zero QCD-axion condensate could indeed develop in NSs. 

We would like to note that our analysis of the physics of a light field such as the QCD-axion in systems at finite chemical potential is of general applicability, therefore of relevance for other light (below the QCD confinement scale) axion-like particles coupled to QCD degrees of freedom, e.g.~\cite{Arvanitaki:2009fg,Graham:2015cka,Nelson:2017cfv,Bellazzini:2017neg,Hook:2019pbh}.~\footnote{A notable scenario is that of the realizations of the relaxion~\cite{Graham:2015cka} which rely on QCD dynamics as the main source of stabilization (back-reaction), see also e.g.~\cite{Nelson:2017cfv}. The potentially dramatic effects of baryon chemical potential and the corresponding implications on NSs for these scenarios will be discussed elsewhere~\cite{Balkin:2021wea}.}
We find this type of new physics well motivated and very interesting, specially in light of the excellent experimental prospects for improving our understanding of dense astrophysical objects such as NSs, as exemplified by the recent detection of gravitational waves from merger events~\cite{TheLIGOScientific:2017qsa}.

The rest of the paper is organized as follows. We review the axion potential in vacuum in~\Sec{mu0section} and how the chemical potential is introduced in quantum field theory, with illustrative examples of boson condensation and two-flavor QCD meson condensation, in~\Sec{ChemFT}. Readers familiar with these topics may skip to \Sec{nucPhase}, where we present our results for the properties of the axion in nuclear matter around nuclear-saturation density. The study of the axion potential in the CFL phase, and our extrapolation to densities expected to be found in NSs cores, is found in~\Sec{CFL}. An overview of the potential effects of an axion condensate in NSs is presented in~\Sec{observablesss}, deferring for future work a more detailed study of this and other implications on NSs and supernovae~\cite{vacNS}. We present our conclusions and outlook in~\Sec{conc}. Finally, several appendices are devoted to detailed calculations of the results presented in the main text.

%%%%%%%%%%%%%%%%%%%%%%%%%%%%%%%%%%%%%%%%%%

\section{Axion potential in vacuum}\label{mu0section}

The most general QCD and axion effective Lagrangian below the electroweak scale, at leading order order in fields and derivatives, is given by
\begin{subequations}
\label{Lqcd}
\begin{align}
\mathcal{L} &=\mathcal{L}_{\text{\tiny QCD}}  +\mathcal{L}_{a }  \,,
\\
\mathcal{L}_{\text{\tiny QCD}} &= -\frac14 G^{\mu\nu}G_{\mu\nu}+ i \bar{q} \slashed{D} q - (\bar{q}_L M q_R + \hc)\,,
\\
\mathcal{L}_{a} &=
\frac12(\partial^\mu a)^2+ 
\left( \frac{a}{f_a} +\theta\right) \frac{ g_s^2}{32\pi^2}G^{\mu\nu}\tilde{G}_{\mu\nu}
+\frac14 a \, g^0_{a\gamma\gamma}F^{\mu\nu}\tilde{F}_{\mu\nu}
+\frac{\partial_\mu a}{2f_a} J^\mu_{\PQ,0}\,, 
\end{align}
\end{subequations}
with implicit flavor and color indices.  $J^\mu_{\PQ} = \sum_q c_q^0\bar{q} \gamma^\mu \gamma_5 q$ is a model-dependent current associated with a spontaneously broken axial $U(1)_{\PQ}$ symmetry, made of the SM matter fields $q$. The Nambu-Goldstone boson (NGB) of the $U(1)_{\PQ}$ is the axion field $a(x)$, with decay constant $f_a$ defined by its coupling to gluons. The axion coupling to photons is given by $g^0_{a\gamma\gamma} = \frac{e^2}{8\pi^2 f_a}\frac{E}{N}$, with $E/N$ the ratio of the electromagnetic (EM) and the color anomalies. 

In the free and chiral limit the theory is invariant under the symmetry group
\begin{align}
SU(3)_c \times SU(N_f)_L \times SU(N_f)_R  \times U(1)_{B} \times U(1)_{A}\,, \label{FullSym}
\end{align}
with the quark representations, for $N_f=3$ (the number of flavors we consider in this work) 
\begin{align}
q_L : ({\bf 3,3,1})_{+1,+1}\,, \;\;\;\; q_R : ({\bf 3,1,3})_{+1,-1}\,.
\end{align}
At low energies, QCD confines and a chiral condensate $\vev{\bar q_R q_L}$ develops that breaks spontaneously the global symmetries
\begin{align}
SU(N_f)_L \times SU(N_f)_R \times U(1)_B\times U(1)_A \to SU(N_f)_{R+L}\times U(1)_B\,.
\end{align}
The low energy degrees of freedom are described by the fluctuations of the condensate, i.e.~the NGBs of the broken chiral symmetries~\footnote{We chose to include the $\eta'$, even though it is not well described as a NGB (unless in the large $N_c$ limit), to make explicit the similarities with the effective Lagrangian in the CFL phase, see~\Sec{CFL}.}
\begin{align}
\Phi \equiv  \exp \left[ \frac{i \pi^a \lambda^a}{f_\pi} \right]\exp \left[ \frac{i \eta'  }{f_{\eta'}N_f} \right]  \equiv \Sigma \exp \left[ \frac{i \eta'  }{f_{\eta'}N_f} \right] \,,
\end{align}
where $\lambda^a$ for $a=1,..,N^2_f-1$ are the $SU(N_f)$ generators with the normalization convention $\text{Tr}[\lambda^a \lambda^b]= 2\delta^{ab}$. Under the symmetries in (\ref{FullSym}), $\Phi$ transforms as
\begin{align}
\Phi : ({\bf 1,3,\bar{3}})_{0,+2}\,.
\end{align}

The explicit breaking of the chiral symmetries by the quark masses can be incorporated in the low-energy theory by promoting the quark mass matrix, $M = \text{Diag}[m_u,m_d,m_s]$, to a spurion with the transformation properties
\begin{align}
M : ({\bf 1,3,\bar{3}})_{0,+2}\,.
\label{massTransformation}
\end{align}
Under a $U(1)$ axial rotation, the $\theta$ angle shifts as $\theta  \to \theta + 2N_f \alpha_A$. The $U(1)_A$ is therefore anomalous, explicitly broken by non-perturbative effects associated with incalculable large instantons. Since the shift symmetry of the axion, associated with $U(1)_{\PQ}$, can be used to remove the $\theta$ angle from the Lagrangian \Eq{Lqcd}, $a \to a - \theta f_a$, the axion can be treated as an actual dynamical spurion for the $U(1)_A$. 

The non-perturbative nature of the axial anomaly means that the effective Lagrangian for the $\eta'$, which shifts under $U(1)_A$ as $\eta' \to \eta' + 2N_f \alpha_A f_{\eta'}$, is not calculable.~\footnote{If a perturbative expansion in the number of instantons were possible, the leading effective potential would read ${V}_{0} = b (\text{Tr}[\Phi^\dagger M]+\hc) - c (e^{-ia/f_a} \text{det} \Phi^\dagger + \hc)$. This will in fact be the case in the CFL phase, see~\Sec{CFL}. \label{footetap}} 
That would be the case for the axion as well, if not for the fact that one can move to a different basis by performing a local chiral transformation of the quarks in \Eq{Lqcd},
\begin{align}
q \to e^{\frac{i a(x)}{2 f_a}\gamma_5Q_a}q\,,
\label{ChiralTrans}
\end{align}
with $Q_a$ an arbitrary matrix in flavor space which, if $\Tr[Q_a] = 1$, eliminates the axion coupling to gluons.
In this basis, the Lagrangian above the QCD confinement scale reads
\begin{subequations}
\label{genericBasis}
\begin{align}
\mathcal{L}_{\text{\tiny QCD}} &= -\frac14 G^{\mu\nu}G_{\mu\nu}+ i \bar{q} \slashed{D} q - (\bar{q}_L M_a q_R + \hc) \,, \;\; M_a  \equiv e^{\frac{ia(x) Q_a}{2f_a}} M e^{\frac{ia(x) Q_a}{2f_a}}\,,
\\
\mathcal{L}_{a} &=
\frac12(\partial^\mu a)^2
+\frac14 a \, g_{a\gamma\gamma}F^{\mu\nu}\tilde{F}_{\mu\nu}
+\frac{\partial_\mu a}{2f_a}J^\mu_{\PQ}
\\
&J^\mu_{\PQ} = \sum_q c_q\,  \bar{q}  \gamma^\mu \gamma_5 q \,, \;\;  c_q \equiv c_q^0-[Q_a]_q\,, \quad g_{a\gamma\gamma} = \frac{e^2}{8\pi^2 f_a}\left( \frac{E}{N}-6 \text{Tr}[Q_aQ_e^2]\right)
\label{DerCouplings}
\end{align}
with $Q_e = \text{Diag}[2/3,-1/3,-1/3]$ the flavor-space matrix of electric charges.
\end{subequations}
After such a redefinition of the quark fields, and upon integrating out the heavy $\eta'$, the axion potential can be related to that of the QCD pions
\begin{align}
{V}_0 = b (\text{Tr}[\Sigma^\dagger M_a]+\hc) \,,
\label{VzeroDensity}
\end{align}
with
\begin{align}
b 
=  -\frac{m_\pi^2 f_\pi^2}{2(m_u+m_d)}\,,
\label{bchiral}
\end{align}
where $m_\pi$ is the neutral pion mass and we neglected $O(\Delta m/m_s)$ terms, $\Delta m \equiv \frac12 (m_u-m_d)$.

In this $N_f=2$ approximation, the quark condensates evaluated in the vacuum are $2b = \vev{\bar{q} q}_0 \equiv\frac12 {\left< \bar{u}u+\bar{d}d \right>_0}$, which leads to the Gell-Mann-Oakes-Renner (GOR) relation
\begin{align}
\vev{\bar{q}q}_0(m_u+m_d)=-m_\pi^2 f_\pi^2\,. 
\label{GOR}
\end{align}
The axion mass can be calculated at leading order by integrating out the chiral NGBs at tree-level, see~\App{app1}.
The final result for the axion mass reads
\begin{align}
&(m_a^2)_0 =  \frac{m_\pi^2 f_\pi^2}{f_a^2}\frac{  m_u  m_d }{ (m_u+m_d)^2} \,, 
\label{axionMassZeroDensity}
\end{align}
where we neglected corrections of order $O(m_{u,d}/{m_s})$, since they are numerically of the same order as other next-to-leading order (NLO) corrections (e.g.~$\eta'$ mixing)~\cite{diCortona:2015ldu}.
We finally note that, as it could not be otherwise, the axion mass is independent of the arbitrarily chosen $Q_a$. Such a matrix however can be chosen to simplify the calculation of a given observable. For example, choosing $Q_a = \frac{M^{-1}}{\text{Tr}[M^{-1}]}$ removes all the tree-level mixing between the axion and the neutral mesons, thus simplifying e.g.~the calculation of the axion mass.

%%%%%%%%%%%%%%%%%%%%%%%%%%%%%%%%%%%%%%%%%%

\section{Chemical potential in quantum field theory} \label{ChemFT}

Introducing a chemical potential in quantum field theory is a generalization of the procedure in statistical mechanics. One defines a new operator corresponding to the thermodynamic Landau free energy (a.k.a.~grand thermodynamic potential density)
\begin{align}
\hat \Omega = \mathcal{H}-\mu_i J_{i}^0 \,,
\end{align}
with $\mathcal{H}$ the Hamiltonian density, $J_{i}^0$ the conserved charge density associated with a given global symmetry of the system (i.e.~the temporal component of the conserved current), and $\mu_i$ the corresponding chemical potential.~\footnote{We recall that the grand-canonical density matrix is given by $\hat \rho = \exp \left[ -\beta (H - \mu_i Q_i) \right]$, with $\beta = 1/T$ ($T$ is the temperature), $H$ the Hamiltonian, and $Q_i$ the conserved charge. The partition function is then $Z(\mathcal{V},T,\mu_i) = \Tr \hat \rho$, where $\mathcal{V}$ is the volume, eventually taken to infinity. The thermodynamic potential density is $\Omega(T,\mu) = -(T/\mathcal{V}) \ln Z = \rho - \mu_i n_i = - p$, with $\rho$ the energy density, $n_i$ the number density, and $p$ the pressure. The grand-canonical average of an operator $\mathcal{O}$ is then $\vev{O}_{T,\mu_i} = \Tr[\mathcal{O} \hat \rho]/Z$ (with a slight abuse of notation, when clear we will denote ensemble averages simply by $\vev{O}$). Then $n_i = \vev{J_i^0} = - (\partial \Omega/\partial \mu)_T$, while the entropy density is given by $s = - (\partial \Omega/\partial T)_\mu$.}
From the path integral representation of the partition function (see e.g.~\cite{Kapusta:2006pm}), one arrives at the following prescription: the temporal derivative of each field transforming under the global symmetry in question is shifted by
\begin{align}
\partial_0 \to \partial_0+ i \mu_i T_i^{\mathcal{R}}\,, \label{chemicalPotental}
\end{align}
with $T_i^{\mathcal{R}}$ the generator of the global symmetry in the appropriate representation $\mathcal{R}$. Chemical potential therefore acts as a source for the temporal component of the corresponding conserved current, much like a background gauge field potential. Since it singles out the time direction, the chemical potential breaks the Lorentz symmetry down to its $SO(3)$ subgroup of spatial rotations. Charge conjugation symmetry (C), under which $J_{i}^0 \to -J_{i}^0$, is also broken, while parity (P) and time-reserval (T) are preserved -- CP and CPT are thus broken. If part of a non-abelian group, a chemical potential also breaks the global symmetry by singling out a specific direction in generator space, namely $\mu_i T_i$, which defines an unbroken $U(1)$ subgroup.

\subsection{$U(1)$ toy model} \label{toy}

A simple toy model that illustrates the main effect of the chemical potential is a complex scalar theory with a global $U(1)$ symmetry~\cite{Haber:1981fg,Kapusta:1981aa}. After using the prescription of \Eq{chemicalPotental}, one finds the following Lagrangian
\begin{align}
\mathcal{L}({\mu}) = \partial_\mu \phi^* \partial^\mu \phi+i\mu(\phi \partial_0 \phi^*-\phi^* \partial_0 \phi) - (m^2-\mu^2)|\phi|^2-\lambda |\phi|^4\,.
\end{align}
For $m^2>\mu^2$, the field expectation value is trivial, $\vev{\phi} = 0$, and respects the global $U(1)$ symmetry. The two propagating degrees of freedom have different dispersion relations 
\begin{align}
\omega_\phi(\vec{k}) = \sqrt{k^2+m^2}-\mu\,, \;\;\;\; \omega_{\phi^*}(\vec{k}) = \sqrt{k^2+m^2}+\mu\,.
\label{dispersiontoy}
\end{align}
The appearance of the chemical potential breaks C symmetry, which appears as a $\phi~\leftrightarrow~\phi^*$ exchange symmetry in the $\mu=0$ theory -- therefore $\mu$ can be treated as a spurion transforming as $\mu \to -\mu$.  

Above the threshold $|\mu| > m$, the global $U(1)$ is spontaneously broken by the expectation value and the theory describes a Bose-Einstein condensate (BEC) phase. In contrast to the \emph{ideal} ($\lambda=0$) ultra-relativisitic Bose gas~\cite{Haber:1981fg}, in the \emph{interacting} theory (with $\lambda>0$) the chemical potential can be larger than $m$~\cite{Kapusta:1981aa}, without leading to any inconsistencies. Note, that our fundamental potential being the Landau free energy $\Omega$, the fixed thermodynamical parameter is $\mu$, which sets the effective energies of the particles in the system due to a coupling to the ``particle bath''. This allows the flow of particles in and out of the system, implying that the charge density, $n_\phi-n_{\phi^*}$, is a derived quantity set by $\mu$.\footnote{This is in complete analogy to temperature $T$, with sets the effective energy of particles in the system due to a coupling to a ``heat bath''. This allows the flow of heat in and out of the system, implying that the entropy of the system is a derived quantity set by $T$.} As we show below, for $|\mu|>m$, the $T=0$ system contains non vanishing charge density in the form of the BEC. One can interpret this appearance of charge as particles from the ``particle bath'' being inserted in the ground state of the system.

In the BEC phase, the following parameterization is useful
\begin{align}
\phi(x) = \frac{1}{\sqrt{2}} e^{i\chi(x)/v}(v+\sigma(x))\,.
\end{align}
The classical potential is minimized for $v^2 = \frac{\mu^2-m^2}{\lambda} $, and we find the following Lagrangian
\begin{subequations}
\begin{align}
\mathcal{L}(\mu) &= \frac12\left[ (\partial_\mu \chi)^2\left(1+\frac{\sigma}{v}\right)^2+(\partial_\mu \sigma)^2\right]+\mu v  \left(1+\frac{\sigma}{v}\right)^2 \partial_0 \chi-V(\mu)\,,
\\
V(\mu) &= \frac12 m_\sigma^2 \sigma^2 +\lambda v\sigma^3+\frac14\lambda \sigma^4-\frac14\lambda v^4\,,
\end{align}
\end{subequations}
with $m_\sigma^2 = 2\lambda v^2 = 2(\mu^2-m^2)$. The charge density in the condensed phase is non-vanishing in the limit of zero temperature $\beta \equiv 1/T \to \infty$ and infinite volume $\mathcal{V}\to \infty$, 
\begin{align}
(n_\phi-n_{\phi^*})|_{T=0} = 
\lim_{\beta,\mathcal{V} \to\infty} \frac{1}{\beta \mathcal{V}} \left(\frac{\partial \ln Z}{\partial \mu}\right)_{\beta} =
-\left(\frac{\partial V}{\partial \mu}\right)\biggr|_{\vev{\sigma}=0}= \frac{\mu^3}{\lambda}\left(1-\frac{m^2}{\mu^2}\right)\,,
\end{align}
where we used the classical ($\hbar\to0$) result for the generating functional $\ln Z = -\beta \mathcal{V} V(\mu)$ for a homogeneous classical configuration $\vev{\sigma}$.
By diagonalizing the quadratic field operators in momentum space one finds the dispersion relations for the two propagating degrees of freedom
\begin{align}
\omega_\pm^2(\vec{k}) = (3\mu^2-m^2)\left[1+\frac{k^2}{3\mu^2-m^2} \pm \sqrt{1+\left(\frac{2 \mu k}{3\mu^2-m^2}\right)^2}\right]\,,
\end{align}
which at zero momentum are
\begin{align}
& \omega_-(\vec{0}) =  0\,, \quad
\omega_+(\vec{0}) =  \sqrt{6\mu^2-2m^2}\,.
\end{align}
As expected, there is one massless excitation, corresponding to the NGB of the spontaneously broken $U(1)$, and one massive excitation, the radial (or Higgs) mode.

\subsection{Meson condensation} \label{MesonCondensation}

We review now the importance of a chemical potential in the context of meson condensation in QCD, in particular for the case of two flavors~\cite{Son:2000xc,Kogut:2001id,Mammarella:2015pxa,Mannarelli:2019hgn}, and discuss for the first time its effects on the axion potential. This is a simplified version of the more complicated, but plausibly more realistic, scenario of kaon condensation ($N_f = 3$), to be discussed in~\Sec{KaonCondensation}. 

For $N_f=2$, the chiral condensate breaking $SU(2)_{L}\times SU(2)_{R}\times U(1)_B \times U(1)_A$ spontaneously to $SU(2)\times U(1)_{B}$ can be parameterized, in full generality, as
\begin{eqnarray}
&\vev{\bar{q}_R q_L}  \equiv \vev{\bar{q}_R q_L}_0 \, e^{i\alpha} \, \Sigma_0 \,,& \nonumber \\
&\Sigma_0 = \cos \theta\, \mathbb{1}_2+i \sin\theta\, \hat{n} \cdot\vec{\sigma}\,, \quad \hat{n} = (\sin\psi\cos\chi,\sin\psi\sin\chi,\cos\psi)\,,& \label{sigma0}
\end{eqnarray}
where $-\pi/2 \leq \theta <\pi/2$,\footnote{The shift $\theta \to \theta+\pi$ can be compensated by shifting $\alpha \to \alpha+\pi$.
%One can check explicitly that the latter shift leaves the Lagrangian invariant in $N_f=2$.
}
and with the sigma field transforming as
\begin{align}
\Sigma_0 \to L \Sigma_0 R^\dagger\,.
\label{sigma}
\end{align}
In \Eq{sigma0}, we used the fact that a field transforming as a bi-fundamental under $SU(2)_L \times SU(2)_R$ can be written as a radial mode, here frozen to some constant value $\vev{\bar{q}_R q_L}_0$, times a 2-by-2 unitary matrix $\Sigma_0$, which parameterizes the orientation of the ensemble average in the presence of finite $\mu$, which we call the orientation of the expectation value here.~\footnote{$SU(2)_L$ and $SU(2)_R$ are generated by $T_L^a = \frac12\sigma^a$ and $T_R^a = \frac12\sigma^a$, respectively, where as usual it should be understood that the $L$ and $R$ operators act on different indices and therefore commute.}

The phase factor $e^{i\alpha}$ is identified with the direction in field space associated with the anomalous axial $U(1)$. A potential for $\alpha$ is generated by non-perturbative effects, whose minimum is at $\alpha=0$, which we take from this point on.
The angles defined in \Eq{sigma0} can be related to expectation values of the usual pion fields (at vanishing chemical potential)
\begin{align}
\theta \equiv \frac{\vev{\Pi}}{f_\pi}\,, \;\; \frac{1}{\sqrt{2}}\sin\psi e^{\mp i\chi} = \frac{\vev{\pi_\pm}}{\vev{\Pi}}\,, \;\; \cos\psi = \frac{\vev{\pi_3}}{\vev{\Pi}}\,,
\end{align}
where we defined $\sqrt{\vev{\pi_i \pi_i}} \equiv \left< \Pi \right>$.
In Dirac notation
\begin{align}
&\vev{\bar{q} q} =  \frac12 \vev{\bar{q} q}_0 (\Sigma_0+\Sigma_0^\dagger)\,,\;\;\; \vev{\bar{q}i \gamma_5 q} = \frac{1}{2i} \vev{\bar{q} q}_0(\Sigma_0-\Sigma_0^\dagger)\,,
\end{align}
where we denoted $ \vev{\bar{q}_R q_L}_0= \vev{\bar{q}_L q_R}_0 \equiv \vev{\bar{q} q}_0/2$.
Therefore CP is broken in the ground state if $\Sigma_0 \neq \Sigma_0^\dagger$, that is if $\theta \neq 0$.

We wish to study this system at a non-vanishing chemical potential for isospin
\begin{align}
\hat{\mu} = \muI(T_L^3+T_R^3)\,,
\label{muDef}
\end{align}
and we shall neglect for the remainder of this section isospin breaking due to the quark masses and electromagnetic interactions, making the choice in \Eq{muDef} completely generic. Such a chemical potential is associated with the $\sigma_3$ rotation of the vector $SU(2)_{L+R}$ subgroup. Therefore, according to \Eqs{chemicalPotental}{sigma}, we promote the temporal derivative of $\Sigma_0$ to
\begin{align}
\partial_0 \Sigma_0 \to \partial_0 \Sigma_0 +i\muI T^3_L  \Sigma_0-i\muI \Sigma_0  T^3_R = 
\frac{i}{2}\muI [\sigma_3 ,\Sigma_0]\,.\label{covDer}
\end{align}
Note that changing $\hat{\mu}  \to \hat{\mu} + \frac16 \mathbb{1}_2$ in \Eq{muDef} has no effect on \Eq{covDer} and on the following derivation, therefore in this context the isospin chemical potential can be equivalently associated with the chemical potential for electric charge.\footnote{One can then think of $\muI$ as a non-vanishing averaged value for the zero component of the photon field $\muI = \vev{A_0}$, which can be intuitively understood as a classical background electric charge density.}
The resulting potential for the pions and the axion, the latter entering via the quark mass matrix, $M = m \mathbb{1}_2$ ($m_u = m_d$), as in \Eq{genericBasis} (with $Q_a = \mathbb{1}_2/2$), is given by
\begin{align}
&V =
\frac{f_\pi^2\muI^2}{16}\text{Tr}[ [\sigma_3 ,\Sigma_0] [\sigma_3 ,\Sigma_0^\dagger]] 
+
\frac{\vev{\bar{q}q}_0}{2 }\text{Tr}[\Sigma_0 M e^{-\frac{ia}{2f_a}}+\Sigma^\dagger_0 M e^{\frac{ia}{2f_a}}]\,, \label{L1}
\end{align}
at leading order in $m/\Lambda_{\chi}$ and $\mu/\Lambda_{\chi}$, $\Lambda_{\chi}$ being the cutoff of the chiral Lagrangian. 
We note that the first term arises from the usual kinetic term, $\frac{1}{4}f_\pi^2 \Tr[\partial_\mu \Sigma_0 \partial^\mu \Sigma_0^\dagger]$, after the replacement (\ref{covDer}). 
Using \Eq{sigma0} we find
\begin{align}
V = -\frac12 \muI^2 f_\pi^2 \sin^2\theta \sin^2\psi - m_\pi^2 f_\pi^2 \cos \theta \cos \left( \frac{a}{2f_a}\right)\,, \label{VthetaPion}
\end{align}
where $m_\pi$ here is the neutral pion mass in vacuum, i.e.~$m_\pi^2 f_\pi^2 = - 2 m \vev{\bar{q}q}_0$.
We see that the isospin chemical potential introduces an additional source of explicit symmetry breaking -- while leaving unbroken the $U(1)_{L+R}$ symmetry defined by the generator in \Eq{muDef}, $\hat{\mu}$ explicitly breaks the shift symmetries associated with the would-be NGBs charged under $U(1)_{L+R}$, i.e. the charged pions. Indeed, as discussed above, $U(1)_{L+R}$ is equivalent to the electric charge. Consequently, the first term in \Eq{VthetaPion} is proportional to the expectation value of the charged pions, $\sin^2 \theta \sin^2 \psi \propto \vev{\pi^+ \pi^-}$. Since $\hat \mu$ commutes with the $U(1)_{L-R}$ associated with the neutral pion, the neutral NGBs are unaffected by the chemical potential  and the potential \Eq{VthetaPion} is minimized at $\vev{\pi_3} = 0$ ($\psi=\pi/2$) and $\vev{a} = 0$ as in the $\mu=0$ vacuum.

The minimum of the potential for any value of $\muI$ is then found at
\begin{align}
&\cos \theta = \text{Min}\left[1, \frac{m_\pi^2}{\muI^2}\right]\,.
\end{align}
For $|\muI|<m_\pi$, the ground state is the trivial one, $\Sigma_0=1$, thus its orientation is the same as for $\muI=0$. For $|\muI|>m_\pi$, pion condensation takes place and the orientation of the expectation value is no longer trivial. We note that in this case $\chi$ constitutes a flat direction which, as we confirm later, corresponds to a NGB from the spontaneous breaking of electric charge, $U(1)_{L+R}$. Setting, without loss of generality, $\chi=0$, we can write the QCD orientation for $|\muI|>m_\pi$ as
\begin{align}
\Sigma_0 = \begin{pmatrix}
\cos \theta & i\sin\theta \\
i\sin\theta & \cos\theta
\end{pmatrix} \,.
\end{align}
At this point we recall that since $\theta \neq 0$, CP is broken by the expectation value, a result of a sufficiently large explicit breaking of CP by the chemical potential in the \emph{charged} pion sector. Instead, CP-invariance in the \emph{neutral} sector is preserved by the charge chemical potential, which leaves the expectation values in that sector untouched. We see now that only if $\vev{\pi_3} \neq 0$ ($\psi \neq \pi/2$) could the axion condense, which requires additionally explicit breaking of isospin, i.e. $m_u \neq m_d$.

Having established the Goldstone boson expectation values at finite-density, let us turn our attention to their fluctuations. Since these are associated with the $SU(2)_L \times SU(2)_R$ generators broken by $\Sigma_0$, we define the following rotated generators
\begin{align}
(T_L^a)_\theta = \xi_0 (T_L^a) \xi_0^\dagger\,, \;\; (T_R^a)_\theta =\xi_0^\dagger (T_R^a) \xi_0\,,
\label{rotatedT}
\end{align}
where $\xi_0 \equiv \sqrt{\Sigma_0}$. The broken and unbroken generators are then given by
\begin{align}
X^a = (T_L^a)_\theta-(T_R^a)_\theta\,, \;\;\; T^a = (T_L^a)_\theta+(T_R^a)_\theta\,,
\end{align}
respectively. The fluctuations around the $\Sigma_0$ ground state can be parameterized as
\begin{align}
\Sigma = \xi_L \Sigma_0 \xi_R^\dagger = 
\exp\left[\frac{i\pi^a (T_L^a)_\theta}{f_\pi} \right]\Sigma_0\exp\left[\frac{i\pi^a (T_R^a)_\theta}{f_\pi} \right] = 
\xi_0 \exp\left[\frac{i\pi^a \sigma^a}{f_\pi} \right] \xi_0\,,
\label{fluc}
\end{align}
where, abusing notation, we have written the (pseudo-)NGBs as $\pi^a$, like the standard $\theta = 0$ pions.~\footnote{We note that, given \Eq{rotatedT} and $T_L^a = \frac12\sigma^a$, $T_R^a = \frac12\sigma^a$, it follows that $\xi_L = \xi_0 \exp \left[\frac{i\pi^a \sigma^a}{2 f_\pi} \right]\xi_0^\dagger$ and $\xi_R = \xi_0^\dagger \exp \left[-\frac{i\pi^a \sigma^a}{2 f_\pi} \right]\xi_0$.}

\begin{figure}
\centering
\includegraphics[width=0.7\textwidth]{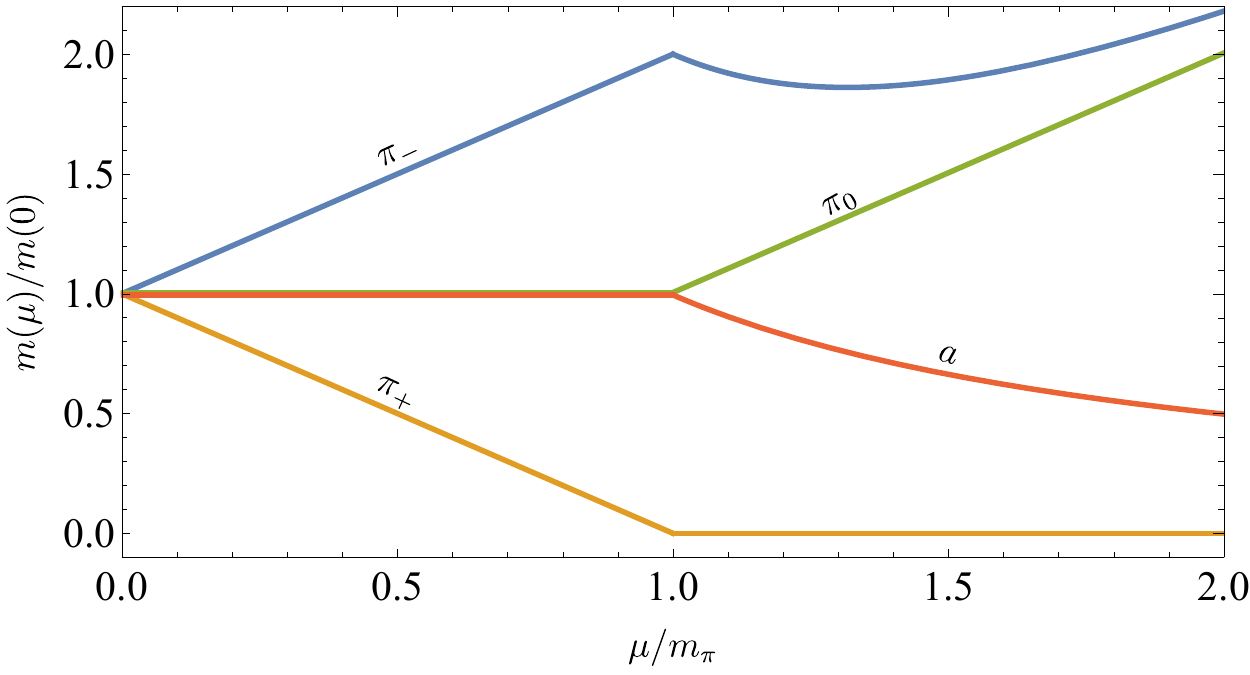}
\caption{Mass spectrum of the vacuum excitations as a function of $\muI/m_\pi$. The masses are normalized to their respective $\muI=0$ value. The charged ${\pi}_+$ and ${\pi}_-$   modes (orange and blue curves respectively) evolve similarly as the $\phi$ and $\phi^*$ modes in the $U(1)$ toy model: a linear split in masses in the uncondensed phase, continuously transitioning to a massless Goldstone mode and a massive radial mode in the condensed phase. The masses of the neutral modes, $\pi_0$ and $a$ (green and red curve respectively) are unaffected by the chemical potential in the uncondensed phase. In the condensed phase, $m_{\pi_0}$ increase linearly with $\muI$, while the axion becomes lighter as $\muI$ increases.}
\label{pionCondensation}
\end{figure}

The dispersion relations for the neutral degrees of freedom, $\pi_0$ and the axion, are the same as for vanishing chemical potential. Their masses can be obtained from \Eq{L1} (with the substitution of $\Sigma_0$ by $\Sigma$),
\begin{align}
(m_{\pi_0}^2)_{\theta} = m_{\pi}^2/\cos\theta \,, \quad (m_a^2)_{\theta} =(m_a^2)_0 \cos\theta \,,
\label{axionMassTheta}
\end{align}
with $(m_a^2)_0$ the mass of the axion in vacuum, \Eq{axionMassZeroDensity}, and where we note that for $|\muI|>m_\pi$, $(m_{\pi_0}^2)_{\theta} = \muI^2$.
The change of the axion mass for $\theta \neq 0$ simply follows from the fact that, once the mixing with $\pi_3$ is eliminated, it has to be proportional to the CP-even combination $\Tr[\Sigma_0+\Sigma_0^\dagger] \propto \cos\theta$.
The increase in the neutral pion mass can be understood as a result of its repulsive interaction with the charged pions. 
The dispersion relation for the charged pions is very similar to the $U(1)$ toy model of~\Sec{toy}.
In the uncondensed phase $|\muI|<m_\pi$, their dispersion relations are
\begin{align}
&\omega_{{\pi}_{\pm}}(\vec{k}) = \sqrt{m_\pi^2+k^2} \mp \muI \,.
\end{align}
Indeed, for the charged states $\pi_{\pm}\equiv \frac{1}{\sqrt{2}}(\pi_1 \mp i \pi_2)$ we recognize the same mass splitting we found in \Eq{dispersiontoy}.
In the condensed phase $|\muI|>m_\pi$, the remaining $U(1)_{L+R}$ symmetry is spontaneously broken. The effective masses of the charged pions are
\begin{align}
&\omega_{{\pi}_{+}}(\vec{0}) = 0 \,, \quad \omega_{{\pi}_{-}}(\vec{0}) = \muI \sqrt{1+\frac{3m_\pi^4}{\muI^4}}\,.
\end{align}
As in the $U(1)$ toy model, the condensed phase contains one massless Goldstone mode and one massive radial mode. In this phase, the system has a non-vanishing charge density
\begin{align}
n_{{\pi}_+}-n_{{\pi}_-} = -\left(\frac{\partial V}{\partial \mu}\right)\biggr|_{\pi_i=a=0} = f_\pi^2 \muI \left(1-\frac{m_\pi^4}{\muI^4}\right)\,.
\end{align}
The effective masses of the pions and the axion are plotted in \Fig{pionCondensation}.

%%%%%%%%%%%%%%%%%%%%%%%%%%%%%%%%%%%%%%%%%%

\section{Nuclear phase} \label{nucPhase}

In this section we study how the properties of the axion, mainly its potential and coupling to nucleons, change in systems at finite baryon density, $n$. In particular, our focus here is on densities around nuclear saturation, $n \sim n_0$, where a description of QCD in terms of hadrons is still meaningful. 

For the axion potential, we identify two main effects: 1) the change in the size and, to some degree, flavor orientation of the quark condensates, as ``measured'' by the mass of the pions (\Sec{qqFiniteDensitySubSection}), and 2) kaon condensation (\Sec{KaonCondensation}), similar to meson condensation, introduced in~\Sec{MesonCondensation}. Both of these effects can be taken into account by a generalization of the axion potential in vacuum, \Eq{VzeroDensity}, to
\begin{align}
V(n) &= \frac12\text{Tr}[\vev{\bar q q}_{n} \hat M_a + \hc  ]\,, \quad 
\hat M_a = \xi_0^\dagger \xi_L^\dagger M_a \xi_R  \xi_0^\dagger \,
\label{PotentialNuclear}
\end{align}
with $M_a$ encoding the dependence on the axion as in \Eq{genericBasis}. $\Sigma_0 = \xi_0^2$ parametrizes the orientation of the QCD ground state that spontaneously breaks $SU(3)_L \times SU(3)_R$ to $SU(3)$ and therefore encodes the effects of kaon condensation. In vacuum, we have $\xi_0 = \mathbb{1}_3$ and the unbroken subgroup is the usual $SU(3)_{L+R}$, while in the kaon-condensed phase, we have $\xi_0 = \xi_0 (\theta)$, with $\theta$ controlling the size of the kaon condensate which, as explained below, ultimately depends on the baryon density. $\xi_{L,R}$ are the 
Goldstone matrices, given by
\begin{align}
&\xi_L = e^{i \frac{\pi^a}{2f_\pi} (T_L^a)_\theta} = \xi_0 \exp\left[\frac{i\pi^a \lambda^a}{2f_\pi} \right] \xi_0^\dagger\, , \quad
\xi_R = e^{-i \frac{\pi^a}{2f_\pi} (T_R^a)_\theta} = \xi_0^\dagger \exp\left[-\frac{i\pi^a \lambda^a}{2f_\pi} \right] \xi_0\,,
\end{align}
a generalization to $SU(3)_L \times SU(3)_R$ of those in \Eq{fluc}.
Finally, the quark condensate $\vev{\bar{q} q}_n$ at finite density becomes a matrix in flavor space,
\begin{align}
\vev{\bar q q}_{n} &= \text{Diag}[\vev{\bar{u}u}_{n},\vev{\bar{d}d}_{n},\vev{\bar{s}s}_{n}]\,,
\end{align}
The detailed derivation of \Eq{PotentialNuclear} is given in \App{appChPT}. The result can also be understood in terms of symmetries: $\hat M_a$ is a spurion that has been dressed by the Goldstones and projected into the $SU(3)_{L+R}$ subgroup. Therefore, it transforms as $\hat M_a \to V \hat M_a V^\dagger$, where $V$ is an $SU(3)_{L+R}$ transformation. $\vev{\bar q q}_{n}$ transforms in the same way, since it is the result of a non-vanishing expectation value of the temporal component of the baryonic current, $n = \vev{J_B^0}$.~\footnote{When the quark condensate is trivial, $\vev{\bar{q} q}_n \propto \mathbb{1}_3$, we recover $V(n) \propto \text{Tr}[\Sigma^\dagger M_a + \hc  ]$. Instead, when the ground state is trivial, $\Sigma_0 = \mathbb{1}_3$, the change in condensates effectively amounts to $m_q \to m_q \vev{\bar{q}q}_n/\vev{\bar{q}q}_0$.}

Concerning the couplings of the axion to nucleons, the main effect we consider can be traced to a change at finite baryonic density of the nuclear matrix elements $\vev{p | \bar{q} \gamma^\mu \gamma^5 q | p}$ (with $p$ the proton, $q = u,d$) as ``measured'' by the axial pion-nucleon coupling (\Sec{axionCouplings}). These set the size of the couplings of protons and neutrons to the axion, as they follow from either its model-dependent UV couplings to light quarks, or from axion-pion mixing. The latter also changes at finite density although, as we explain below, the effect is within known uncertainties.

Before going into the details, several general comments about our treatment of the nuclear medium are in order. To describe the state of the system, we will work directly in terms of baryon densities, $n_p$ and $n_n$ considering only protons and neutrons, respectively. In practice, our independent parameters are the total baryonic density, $n = n_p + n_n$, and the proton fraction, $n_p/n$. This will be more convenient than introducing the corresponding chemical potentials, because our analysis is limited to linear order in $n$, i.e.~we work in the mean-field or Hartree approximation, where e.g.~$n_p \approx \vev{\bar p \gamma_0 p}_{T,\mu_i}$ (and in fact $n_p \approx \vev{\bar p p}_{T,\mu_i}$ in the non-relativistic limit) -- higher-order corrections generically being beyond perturbative control when relevant. Besides, the relative fraction of protons and neutrons is, as shown below, relevant only in our discussion of kaon condensation. There, the chemical potential for electric charge, $\mu$, will also be required to properly describe the system, along with the condition of charge neutrality.
 
\subsection{Quark condensates} \label{qqFiniteDensitySubSection}

We first discuss how the quark condensate changes at finite baryonic density, since this is the most robust effect from the point of view of perturbative control. We derive the implications for the axion mass, which were first noted in~\cite{Hook:2017psm}. The change with density of the quark condensates can be calculated utilizing the Hellmann-Feynman theorem~\cite{Cohen:1991nk}
\begin{align}
\zeta_{\bar{q}q}(n) \equiv \frac{{\vev{\bar{q} q}_{n}}}{\vev{\bar{q} q}_0} = 1 + \frac{1}{ \vev{\bar{q} q}_0} \frac{\partial \Delta E(n)}{\partial m_q}\,, \quad q = u,d,s \,.
\label{qqbarShift}
\end{align}
$\Delta E(n)$ is the energy shift of the QCD ground state due to the finite density background, such that $\Delta E(0)=0$. It can be decomposed as
\begin{align}
\Delta E = E^{\text{free}}+E^{\text{int.}}\,,
\label{deltaE1}
\end{align}
where the first term represents the energy shift due to the presence of a non-interacting Fermi gas, while the second term encodes the energy shift due to nuclear interactions. Neglecting these interactions as well as relativistic corrections, we have $\Delta E = \sum_{x = n,p,\dots} m_x n_x $, and we arrive at the so-called linear approximation for the in-medium condensate
\begin{align}
\zeta_{\bar{q} q }(n) =1+\frac{1}{\vev{\bar{q} q}_{0} }  \sum_x n_x \frac{\partial m_x}{\partial m_q}\,, \quad q = u,d,s\,.
\label{linearApprox}
\end{align}
The derivatives $\partial m_x/\partial m_q$ describe the shift in the nucleon mass due to the non-vanishing quark masses. For two nucleons $\{n,p\}$ and three quarks $\{u,d,s\}$, one naively counts six independent shifts. However, due to the $\{p,u\}\leftrightarrow \{n,d\} $ exchange symmetry, only three shifts are independent. Working in the isospin basis for the quark masses, $\bar{m} \equiv \frac12(m_u+m_d)$ and $\Delta m \equiv \frac12 (m_u-m_d)$, the following sigma terms are identified and defined
\begin{align}
{\sigma_{\pi N}}&\equiv \bar{m}\left( \frac{\partial m_p}{\partial \bar{m}}\right) = \bar{m}\left( \frac{\partial m_n}{\partial \bar{m}}\right)\,, 
\\
{\tilde{\sigma}_{\pi N}} &\equiv \Delta{m}\left( \frac{\partial m_n}{\partial \Delta{m}}\right) = -\Delta{m}\left( \frac{\partial m_p}{\partial \Delta{m}}\right)  \,,
\\
{\sigma_s}&\equiv m_s\left( \frac{\partial m_p}{\partial m_s}\right) = m_s\left( \frac{\partial m_n}{\partial m_s}\right)  \,,
\end{align}
such that
\begin{align}
m_n = M_B + \sigma_{\pi N}+\tilde{\sigma}_{\pi N}+\sigma_s\,, \label{mnDef}
\\
m_p = M_B + \sigma_{\pi N}-\tilde{\sigma}_{\pi N}+\sigma_s\,. \label{mpDef}
\end{align}
with $M_B$ the baryon mass in the chiral limit, $m_q \to 0$. We note that the sigma terms can be expressed in terms of the parameters of the $N_f=3$ chiral Lagrangian for baryons, see \Eq{sigmaParameters} in \App{appChPT}.
The $\sigma_{\pi N}$ and $\sigma_s$ terms have been extracted from pion-nucleon and kaon-nucleon scattering experiments, as well as from lattice simulations by calculating the mass shifts of the nucleons. There are ongoing efforts aimed at the determination of the precise values of these sigma terms. A summary of latest results~\cite{Gubler:2018ctz} shows that their current values are scattered over a fairly wide range, with some tension between experimental and lattice results. In this work we use the conservative estimates $\sigma_{\pi N} = 45\pm15 \MeV$ and $\sigma_s = 30 \MeV$\,. The other sigma term is extracted from the $p-n$ non-electromagnetic mass splitting $2\tilde{\sigma}_{\pi N} = (m_n-m_p)^{\text{non-EM}} = 2\pm0.3 \MeV$~\cite{Gasser:1983yg}. 
Using the GOR relation in \Eq{GOR}, we rewrite the ratios ${\vev{\bar{q} q}_{n}}/\vev{\bar{q} q}_0$ as
\begin{subequations}
\label{qqs}
\begin{align}
\zeta_{\bar{u}u}(n) &= 1-b_1 \frac{n}{n_0} +b_2 \left[2 \frac{n_p}{n}-1\right] \frac{n}{n_0}  \,,
\\
\zeta_{\bar{d}d}(n) &= 1-b_1 \frac{n}{n_0}  -b_2 \left[2 \frac{n_p}{n}-1\right] \frac{n}{n_0}  \,,
\\
\zeta_{\bar{s}s}(n) &= 1-b_3 \frac{n}{n_0} \,,
\end{align}
\end{subequations}
with
\begin{subequations}
\label{c1c2c3}
\begin{align}
b_1 &\equiv \frac{\sigma_{\pi N}n_0}{m_\pi^2 f_\pi^2} =  3.5\times10^{-1}\left( \frac{\sigma_{\pi N}}{45~\text{MeV}}\right)\,,
\\
b_2 &\equiv \frac{\tilde{\sigma}_{\pi N}n_0}{m_\pi^2 f_\pi^2} \frac{\bar{m}}{\Delta{m}}  =  -2.2\times10^{-2} \left(\frac{\tilde{\sigma}_{\pi N}}{1~\text{MeV}} \right) \,,
\\
b_3 &\equiv \frac{\sigma_s n_0}{m_\pi^2 f_\pi^2} \frac{2\bar{m}}{m_s} =   1.7\times10^{-2}\left(\frac{\sigma_s}{30~\text{MeV}} \right)\,.
\end{align}
\end{subequations}
Clearly the $\vev{\bar{s} s}_{n}$ condensate is only weakly affected by the nucleonic background. Therefore, as in vacuum, its contribution to the axion mass will be subleading, being suppressed by $m_{u,d}/m_s$. Additionally, $\vev{\bar{u} u}_{n} \approx \vev{\bar{d} d}_{n}$ up to the small isospin breaking correction~\cite{Meissner:2001gz}, which we neglect. From \Eq{PotentialNuclear} with $\xi_0 = \mathbb{1}_3$ and after taking care of axion-pion mixing (which we discuss in the context of the axion couplings \Sec{axionCouplings}) we reproduce the axion mass at finite density found in~\cite{Hook:2017psm}
\begin{align}
(m_a)^2_{n} = \frac{m_\pi^2 f_\pi^2}{f_a^2}\frac{  m_u  m_d }{ m_u+m_d}\vev{\bar{u}u}_{n}  \approx   (m_a)^2_0 \left(1-b_1 \,\frac{n}{n_0} \right)\,, 
\end{align}
where $m_\pi$ here is the neutral pion mass in vacuum, \Eq{GOR}. In this regard, we note that at the linear order in density the same correction as the axion enters the neutral pion mass in medium, i.e.~$(m_\pi)^2_{n} = m_\pi^2 \vev{\bar{u}u}_{n}$. This is why for the remainder of this section, we shall only consider $n<n_c \equiv n_0/b_1 \approx 2.8\,n_0 \, (45 \MeV/\sigma_{\pi N})$, with $n_c$ being the critical density in which one naively expects chiral symmetry restoration in the linear approximation.

At this point, let us turn our attention to the corrections to the linear, non-relativistic approximation we have considered. This will allow us to estimate the densities up to which our leading result is under perturbative control and can therefore be trusted. First, the energy of a degenerate (zero temperature) ideal Fermi gas receives relativistic corrections. In the fully relativistic limit, the free part of the energy for a fermion $x$ is given by
\begin{align}
E^{\text{free}}_x & = 2\int^{k^x_f} \frac{\mathrm{d}^3k}{(2\pi)^3}\sqrt{k^2+m_x^2}  = m_x n_x F(k_f^x/m_x) \,,  \label{deltaE2}
\\
F(q) &= \frac{3 q \sqrt{q^2+1} \left(2 q^2+1\right)-3 \sinh ^{-1}(q)}{8 q^3} = 1+\frac{3 q^2}{10} + O(q^4) \,,
\end{align}
where $k^x_f$ is the Fermi momentum, $k^x_f = \sqrt{m_x^2 - \mu_x^2}$, which determines the number density,
\begin{align}
n_x = 2\int^{|\vec{k}|\leq k^x_f} \frac{\mathrm{d}^3k}{(2\pi)^3} = \frac{(k^x_f)^3}{3\pi^2}\,.
\label{Fermidensity}
\end{align} 
Therefore, relativistic corrections, of $O((k^x_f/m_x)^2)$, become important at large densities. When this happens, corrections to the QCD ground state energy \Eq{deltaE1} from nucleon interactions become important as well. These are predominantly due to pion exchange, but also from four-baryon contact interactions. It is clear that the latter become important when $n_x/\Lambda_\chi f_\pi^2$ becomes order one. Given \Eq{Fermidensity}, this is also the place where ChPT is beyond control, $k^x_f \sim \Lambda_\chi$, e.g.~the pion-exchange contribution to the energy is not predictable. In addition, since the cutoff of ChPT $\Lambda_\chi$ is numerically close to $m_p \approx m_n$, relativistic corrections are approximately controlled by the same expansion parameter, 
\begin{align}
\frac{k_f^2}{\Lambda_\chi^2} \approx \frac{(3 \pi^2 n/2)^{2/3}}{\Lambda_\chi} \approx (15 \, \%) \, 
\left(\frac{n}{n_0}\right)^{2/3} \, \left(\frac{700 \MeV}{\Lambda_\chi} \right)^2\,,
\end{align}
where we took $k_f = k_f^p \sim k_f^n$. Ultimately the best way to asses the validity of our linear approximation is to explicitly compute the relevant NLO corrections. The interaction energy $E^{\text{int.}}$ has been calculated by summing the so-called Hugenholtz diagrams, which are connected bubble diagrams describing ground-state to ground-state transitions~\cite{Kaiser:2001jx}. The resulting higher-order finite density effects on the quark condensates have been obtained in ChPT for symmetric nuclear matter~\cite{Kaiser:2007nv,Goda:2013bka} and pure neutron matter~\cite{Kaiser:2008qu,Kruger:2013iza}.
These authors have indeed found $O(1)$ deviations from the linear approximation for densities somewhat above nuclear saturation. Specifically, nucleon interactions seem to ameliorate the linear decrease of $\vev{\bar{u} u}_n \approx \vev{\bar{d} d}_n$ in \Eq{qqs}, such that already at $n \approx 2n_0$, the condensates are only at approximately $60\, \%$ of their vacuum value, as opposed to the $15\, \%$ predicted by the linear approximation, and in fact start increasing with density~\cite{Kaiser:2007nv}. This then implies that a more realistic prediction of the axion mass in dense symmetric nuclear matter is 
\begin{align}
(m_a)^2_{n \lesssim 2 n_0} \gtrsim 0.6 \, (m_a)^2_{0}\,.
\end{align}
while for larger densities $n \gtrsim 2_0 \sim n_c$ it becomes difficult to trust the results of ChPT. 

Therefore, the determination of the quark condensates and the axion mass at densities significantly beyond nuclear saturation remains an open and difficult theoretical problem. Importantly, realistic lattice simulations at finite density are currently not feasible due to so-called sign problem. In addition, at such high densities other issues arise (ultimately related to the problem of perturbativity), such as the ``hyperon puzzle'', which concerns the appearance, or absence, of hyperons, see e.g.~\cite{Weise:2019mou} and references therein. In the next section we will focus our attention instead on another effect of strangeness, potentially much more relevant for the fate of the axion at finite density.

\subsection{Kaon condensation} \label{KaonCondensation}

In the previous section we assumed that the vacuum of QCD is trivially oriented and CP-preserving, or equivalently that none of the mesons acquire a non-trivial expectation value. This might, however, not be the case in dense matter. It has been hypothesized~\cite{Kaplan:1987sc} that above certain baryonic densities it becomes energetically possible for the strangeness changing process of a neutron splitting into a proton and a scalar $K^-$ meson, and vice versa, to take place
\begin{align}
n \leftrightarrow p^+ + K^-\,, \label{nDecay}
\end{align}
The reason being the low in-medium kaon mass, which eventually leads to the formation of a $K^-$ condensate. This process takes place along with, and even becomes favored over, the usual neutron $\beta$-decay, $n \to p^++e^-+\bar \nu_e$, and inverse $\beta$-decay, $p^++e^- \to n + \nu_e$, because of the high price of occupying the increasingly energetic Fermi surface of the electrons. Because of this fact, also the processes $e^- \leftrightarrow K^- + \nu_e$ and $e^- \leftrightarrow \mu^- + \bar \nu_\mu + \nu_e$, $\mu^- \leftrightarrow e^- + \nu_\mu + \bar \nu_e$ reach $\beta$-equilibrium~\cite{Thorsson:1993bu}. On the other hand, the formation of a pion ($\pi^-$) condensate ($n \leftrightarrow p^+ + \pi^-$) seems to be disfavored, as we shall discuss below. 

Motivated by these arguments, we shall now entertain the possibility of kaon condensation and derive its effects on the axion potential. Several important comments and some caveats are however in order. We consider this scenario because of the thrilling possibility of leading to axion condensation, even though it takes place -- if it takes place at all -- at densities where a perturbative expansion is questionable, $n \gtrsim 2 n_0$. Because of the inherent uncertainties at such densities, our conclusions will be \emph{qualitative} rather than quantitive. Indeed, similar to our discussion at the end of the previous section on the quark condensates and their finite density corrections beyond the linear approximation, kaon condensation cannot be simply described by the leading order terms in ChPT. In particular, nucleon self-interactions and interactions with pions need to be considered in order to capture the full complexity of this strongly interacting system~\cite{Ramos:2000dq} -- for instance, the latter are the reason behind the fact that $K^-$ condensation is more likely than $\pi^-$ condensation. 
Our working assumption is that all the processes above (neglecting the pions) are in equilibrium, which implies a set of equations relating the chemical potentials of the particle species involved,
\begin{align}
\mu_\mu = \mu_e = \mu_{K^-} = \mu \, , \quad \mu_p - \mu_n = \mu \, ,
\end{align}
where $\mu$ is the chemical potential associated with (positive) electric charge. For convenience, we work directly with muon and electron densities, $n_\mu$ and $n_e$ respectively, both of which are determined by $\mu$ as they follow from an ideal Fermi gas. The size of the kaon condensate, $\theta$, is determined, as in the simple example of meson condensation discussed in~\Sec{MesonCondensation}, by the minimization of the scalar potential, which of course also  determines if the axion condenses or not. Finally, due to the importance of nuclear interactions, the densities of protons and neutron, or equivalently the total baryon density $n$ and the proton fraction $n_p/n$, are not determined by $\mu$. Instead, we enforce the condition of (electric) charge neutrality $n_{\text{EM}} = 0$, and present our results in terms of $n$ and $n_p/n$.\\

An additional important final comment regards the implications of kaon condensation on the NS equation of state (EoS). It has been argued that the inclusion of kaon condensation generically leads to a softer EoS~\cite{Thorsson:1993bu,Glendenning:1997ak}, which usually cannot sustain a large NS mass. This is in conflict with the most massive NSs observed, with masses around $2M_{\odot}$~\cite{Antoniadis:2013pzd,Cromartie:2019kug}. This is the main reason why kaon condensation is currently considered an ``exotic'' possibility. However, axion effects can in fact harden the EoS~\cite{vacNS} and reopen this window. Also, kaon condensation is in fact related to another issue, namely the hyperon puzzle~\cite{Djapo:2008au}. The appearance of hyperons also tends to soften the EoS, resulting in a similar apparent conflict with the observation of massive NSs. Therefore, although the appearance of strangeness seems to be in tension with observations, we think it would be premature to definitively exclude the possibility of kaon condensation at this point, especially in the presence of new physics.\\

\begin{figure}
    \centering
    \includegraphics[width=0.47\textwidth]{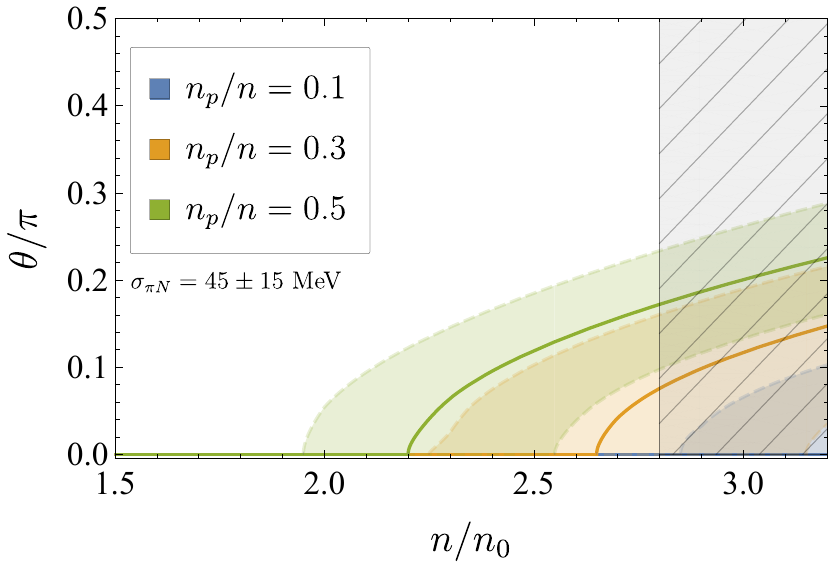}
    \hspace{0.5cm}
    \includegraphics[width=0.47\textwidth]{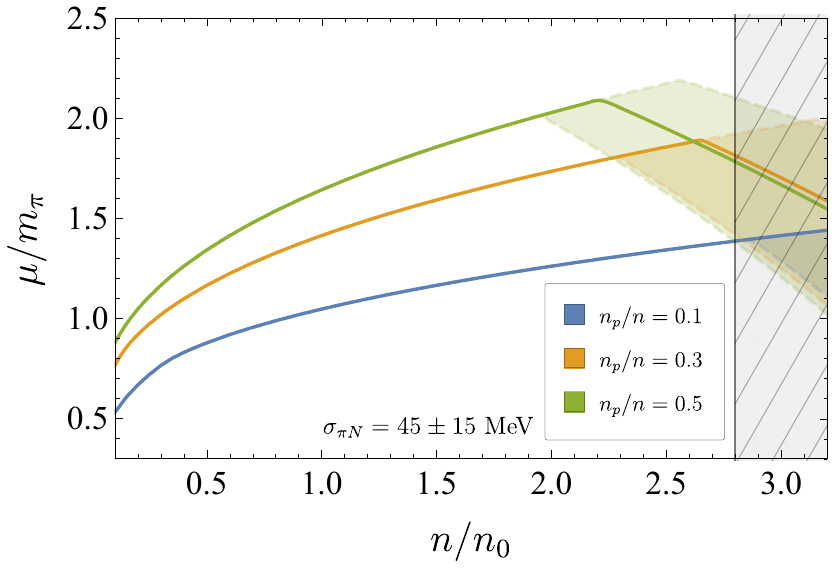}
    \caption{The ground state orientation angle $\theta$ (left panel) and the chemical potential $\mu$ in units of pion mass (right panel) as function of baryon density $n$ for fixed values of proton fraction $n_p/n$. The blue, orange and green curves correspond to $n_p/n=0.1,\,0.3$ and $0.5$, respectively. The solid curves correspond to the numerical solution using the central value of $\sigma_{\pi N} = 45\pm15 \MeV$, while the bands are obtained by the corresponding 1$\sigma$ variation. The gray slashed region corresponds to $n > n_c \approx 2.8\,n_0$ where the quark condensate $\vev{\bar{u} u}_n$ changes sign (for the central value of $\sigma_{\pi N}$).}
    \label{muandthetaPlots}
\end{figure}

\begin{figure}[t!]
    \centering
    \includegraphics[width=0.45\textwidth]{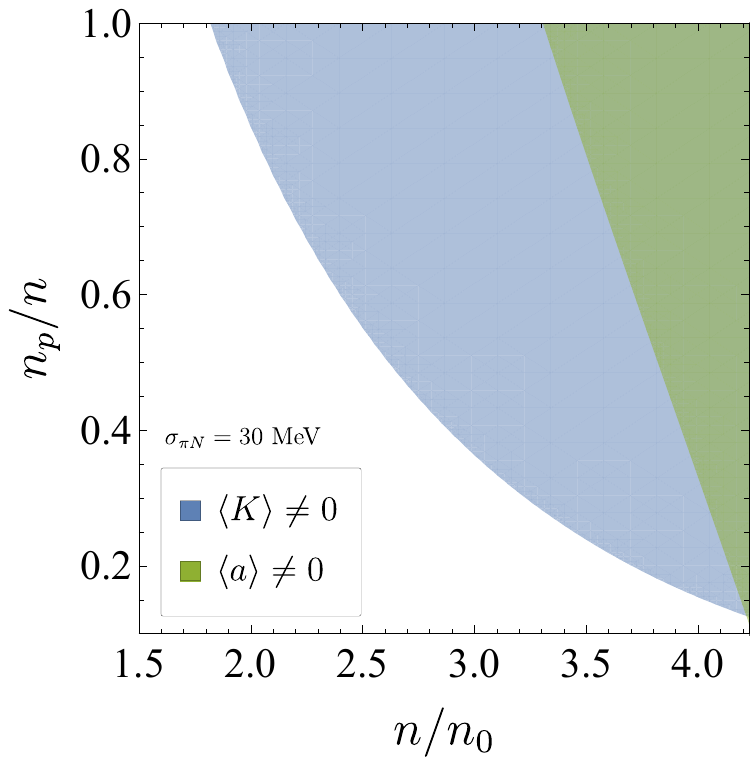}
    \hspace{0.5cm}
    \includegraphics[width=0.45\textwidth]{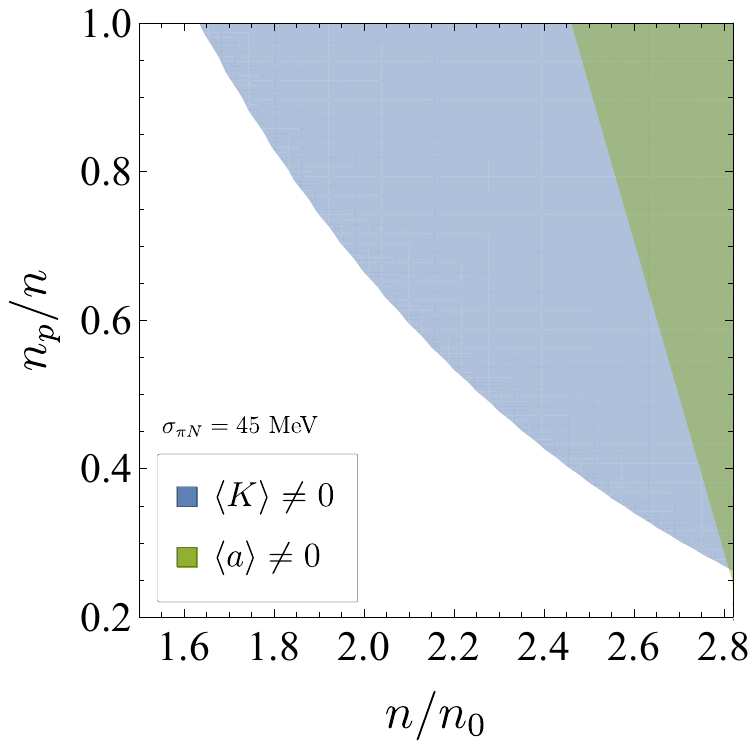}
    \caption{Phase diagram in the plane of baryon density $n/n_0$ and proton fraction $n_p/n$ based on the numerical solution of \Eqs{cosThetaEq}{chargeDensityEq} for $\sigma_{\pi N} = 30 \MeV$ (left panel) and $\sigma_{\pi N} = 45 \MeV$ (right panel). The blue region marks the kaon-condensed phase, while the green region marks the axion-condensed phase. We consider densities in the region $n < n_c$ for the corresponding values of $\sigma_{\pi N}$.}
     \label{phasePlot}
\end{figure}

Let us consider then the possibility that kaon condensation occurs in nuclear matter and qualitatively examine its effects on the axion potential. Once the chemical potential for electric charge $\mu$ is introduced, the dispersion relations for the $K^{\pm}$ modes are given by
\begin{align}
\omega_{K^\pm}(\vec{k}) = \sqrt{\left(m_{K^{\pm}}^2\right)_{n} + k^2}\pm \mu\,,
\end{align}
with the kaon effective in-medium mass
\begin{align}
\left(m_{K^{\pm}}^2\right)_{n} = \frac{1}{f_\pi^2}\left(-\frac{\vev{\bar{u} u +\bar{s} s}_{n}  }{2 } m_s-\frac12 (n + n_p)\, \mu\right)\,. \label{KaonEffectiveMass}
\end{align}
The first term in \Eq{KaonEffectiveMass} is the usual kaon mass to leading order in $m_s$, with the inclusion of the finite density corrections to the relevant quark condensate, which in the linear approximation are given by
\begin{align}
-\frac{\vev{\bar{u} u +\bar{s} s}_{n}  }{2 f_\pi^2 } \, m_s= m_K^2 \,\left(1-\frac12\left[b_1  -b_2 \left(\frac{2\, n_p}{n}-1\right) + b_3 \right] \frac{n}{n_0}\right)\,,
\end{align}
where $m_K^2 = -m_s\vev{\bar{q} q}_0/f_\pi^2$, the neutral kaon mass in vacuum, neglecting $O(m_{u,d}/m_s)$ terms.
The second term in \Eq{KaonEffectiveMass} is a mass correction induced by the baryonic background, due to the model-independent s-wave interactions of the baryons with the mesons, arising from the baryon kinetic term,
\begin{align}
(\mathcal{L}_B)_n = i \Tr[ \bar B \gamma^\mu D_\mu B ] \! \supset \! - \mu \Tr [ \bar B \gamma^0 [\hat Q_e,B] ] \,, \, \hat Q_e = \tfrac{1}{2} \left( \xi_0^\dagger  \xi_L^\dagger Q_e \xi_L \xi_0 + \xi_0  \xi_R^\dagger Q_e \xi_R \xi_0^\dagger \right)
\label{LBdensity}
\end{align}
as it follows from the covariant derivative of ChPT, $D_\mu B = \partial_\mu B + [e_\mu,B]$ with the chiral connection $e_\mu = \frac12 (\xi_0^\dagger  \xi_L^\dagger \partial_\mu \xi_L \xi_0 + \xi_0  \xi_R^\dagger \partial_\mu \xi_R \xi_0^\dagger)$, upon introducing the charge chemical potential, $\partial_0 \to \partial_0 + i \mu Q_e$ with $Q_e = \text{Diag}[2/3,-1/3,-1.3]$, see \App{appChPT} for the details. Note that since $b_2 \ll b_1$, the effective kaon mass decreases with density, and condensation is expected to occur when~\footnote{It is illustrative to also consider the pion effective in-medium mass,
\begin{align}
\left(m_{\pi^{\pm}}^2\right)_{n} = \frac{1}{f_\pi^2}\left(-\vev{\bar{u} u +\bar{d} d}_{n}  \bar{m}+\frac12 (n-2n_p) \mu\right)\,, \label{PionEffectiveMass}
\end{align}
since it shows that, due to the second term and contrary to the kaon, the charge pion becomes heavier with increasing density, at least for a neutron rich background $n_p/n<1/2$~\cite{Muto:1992np}. The argument against pion condensation becomes even stronger when considering higher order terms in ChPT~\cite{Meissner:2001gz}, as we discussed at the end of~\Sec{qqFiniteDensitySubSection} -- these additional corrections even make the pion mass increase with density for $n \approx 2n_0$, even in symmetric nuclear matter, $n_p/n = 1/2$.
}
\begin{align}
\omega_{K^-}(0) = \left({m_{K^\pm}}\right)_{n}-\mu=0
\end{align}
Kaon condensation is introduced by allowing the kaon field to take a non-trivial average value, $\vev{\sqrt{2} K^{\pm}/f_\pi}$, or equivalently, in our notation, reorienting the QCD ground state in medium~\cite{Thorsson:1993bu},
\begin{align}
\Sigma_0 = \begin{pmatrix}
\cos \theta & 0 & i \sin\theta \\
0 & 1 & 0 \\
i\sin\theta & 0 & \cos\theta
\end{pmatrix}\,.
\end{align}
The ground state orientation is determined by the static Lagrangian after setting all the fluctuations to zero. Neglecting for the time being the axion, we find a similar potential to \Eq{VthetaPion}
\begin{align}
V(\theta) =  -\frac12 \mu^2 f_\pi^2 \sin^2\theta-f_\pi^2 (m_{K^\pm}^2)_{n} \cos \theta\,.
\label{Vtheta}
\end{align}
Minimizing $V(\theta)$ leads to the condition
\begin{align}
\cos\theta = \text{Min}\left[1,\frac{(m_{K^\pm}^2)_{n}}{\mu^2}\right]\,, \label{cosThetaEq}
\end{align}
which can be used to determined $\theta = \theta(\mu, n, n_p/n)$.
The requirement of electrical neutrality, $n_{\text{EM}} = - \vev{\partial \mathcal{L}/\partial \mu} = 0$, leads to
\begin{align}
-f_\pi^2\mu \sin^2\theta + \cos \theta \, n_p-\sin^2 \left(\theta/2\right) n_n-n_e(\mu)-n_\mu(\mu) =0\,, \label{chargeDensityEq}
\end{align}
where we included the lepton charge densities, given by
\begin{align}
n_l(\mu) =\Theta(|\mu|-m_l)\, \text{Sign}(\mu) \,\frac{(\mu^2-m_l^2)^{3/2}}{3\pi^2}\,, \;\;\; l=e,\mu\,.
\end{align}
Solving (numerically) the coupled \Eqs{cosThetaEq}{chargeDensityEq}, one can determine the values of $\{\theta,\mu\}$ as a function of $\{n,n_p/n\}$. In \Fig{muandthetaPlots} we show the results for $\theta$ and $\mu$ as a function of baryon density for different values of the proton fraction, while in~\Fig{phasePlot} we plot the region (blue) in the $\{n,n_p/n\}$ plane where $\theta \neq 0$, namely where the system is in the kaon-condensed phase. The evolution for given $\{n,n_p/n\}$ can be understood as follows: for a fixed proton fraction $n_p/n$, as $n$ increases the amount of positive charge due to the protons increases as well, and more leptons are required to satisfy the neutrality condition, leading to an increase in $\mu$. This increase in $\mu$ drives the effective mass of the kaon, \Eq{KaonEffectiveMass}, further down (on top of the decrease in $\vev{\bar{u}u}$ at finite density), until eventually the threshold condition for kaon condensation is met, $\mu= (m_{K^\pm})_{n}$, and a further increase in the proton density can be compensated by inserting $K^-$ particles in the ground state.

Even though we keep them undetermined, let us briefly comment at this point on how the proton fraction could be determined in terms of the total density. Since the interaction energy of nuclear matter also depends on the proton fraction, $E^{\text{int.}} = n \, \epsilon^{\text{int.}}(n,n_p)$, one could enforce that the total energy density is minimal with respect to $n_p/n$, which would lead to the constraint~\cite{Thorsson:1993bu},
\begin{align}
&c_4 \sin^2\left(\theta/2\right)+\mu \cos^2\left(\theta/2\right)    +    \frac{\partial  \epsilon^{\text{int.}}(n,n_p)}{\partial (n_p/n)}=0\,, 
\end{align}
where $c_4 \equiv (2b_2 f_\pi^2 m_K^2)/n_0 \sim 49 \MeV$.

After determining the ground state orientation, let us examine the consequences on the axion potential. The pseudo-NGB potential, after reintroducing the fluctuations we are mainly interested in, namely the neutral mesons $\pi_0,\eta$ and the axion, is given by
\begin{align}
&V(\pi_0,\eta,a) =
\frac{f_\pi^2\muI^2}{4}\text{Tr}[ [Q_e ,\Sigma] [Q_e ,\Sigma^\dagger]]
+\frac12\text{Tr}[\vev{\bar q q}_{n} \hat M_a + \hc  ]
\end{align}
with $\hat M_a$ given in \Eq{PotentialNuclear} and 
\begin{align}
\Sigma = \xi_L \Sigma_0 \xi_R^\dagger = \xi_0 \exp\left[\frac{i\pi^a \lambda^a}{f_\pi} \right] \xi_0 \,.
\end{align}
Note this potential is similar to that discussed in~\Sec{MesonCondensation} in the context of meson condensation, with the additional relevant feature of the density dependent quark condensates, in particular their decrease with density, $\zeta_{\bar{u}u} \approx \zeta_{\bar{d}d}\approx 1-b_1 (n/n_0)$.
The three mass eigenstates, corresponding to mixtures of $\pi_3$, $\eta$ and $a$, have the following masses in the isospin symmetric limit $\Delta m = 0$ and at leading order in $\theta$ and $\bar{m}/m_s$,~\footnote{The calculation of the axion mass is simplified by choosing a particular $\theta$-dependent $Q_a$ matrix which removes the tree-level mixing between the axion and $\pi_0$ and $\eta$, see~\App{axionAnalyticalMass} for more details.}
\begin{align}
&(m^2_{\pi_0})_{n,\theta} \approx m^2_\pi \, \zeta_{\bar{u}u} \left[1+\frac{1}{8}\left(\frac{2\mu^2}{m^2_\pi \zeta_{\bar{u}u}}-\frac{m_s}{\bar{m}}\right)\theta^2 \right]\,, \label{pionmass}
\\
&(m^2_{\eta})_{n,\theta} \approx m_{\eta}^2 \left[1- \frac14 \left(1 + \frac{\zeta_{\bar{u}u}}{4} \right)\theta^2\right]\,,\label{etamass}
\\
&(m^2_{a})_{n,\theta} \approx (m_{a}^2)_0 \, \zeta_{\bar{u}u} \left[1-\frac18 \left(1+\frac{1}{\zeta_{\bar{u}u}}\right)\theta^2\right]\,. \label{axionmass}
\end{align}
with $m_\pi$ and $m_\eta$ the masses in vacuum, respectively \Eq{GOR} and $m_\eta^2 = -4m_s\vev{\bar{q} q}_0/3f_\pi^2$ neglecting $O(m_{u,d}/m_s)$ terms.
\begin{figure}[t!]
    \centering
    \includegraphics[width=0.47\textwidth]{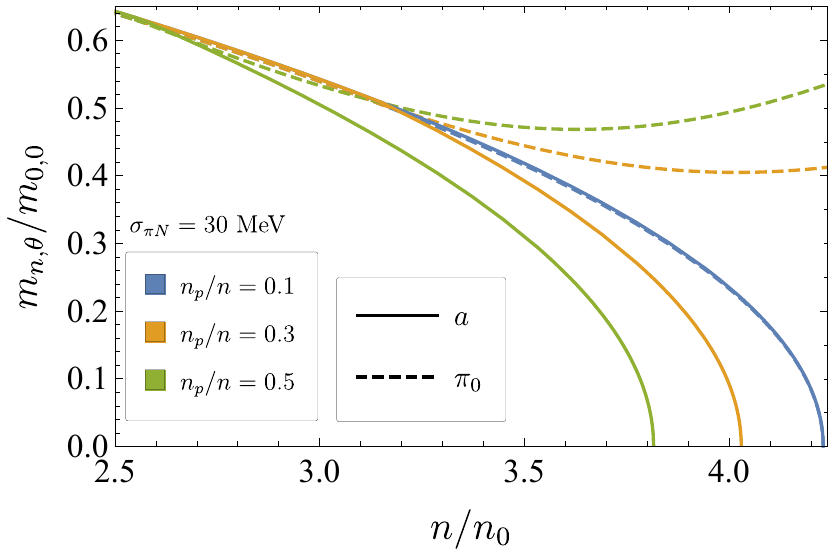}
    \hspace{0.5cm}
    \includegraphics[width=0.47\textwidth]{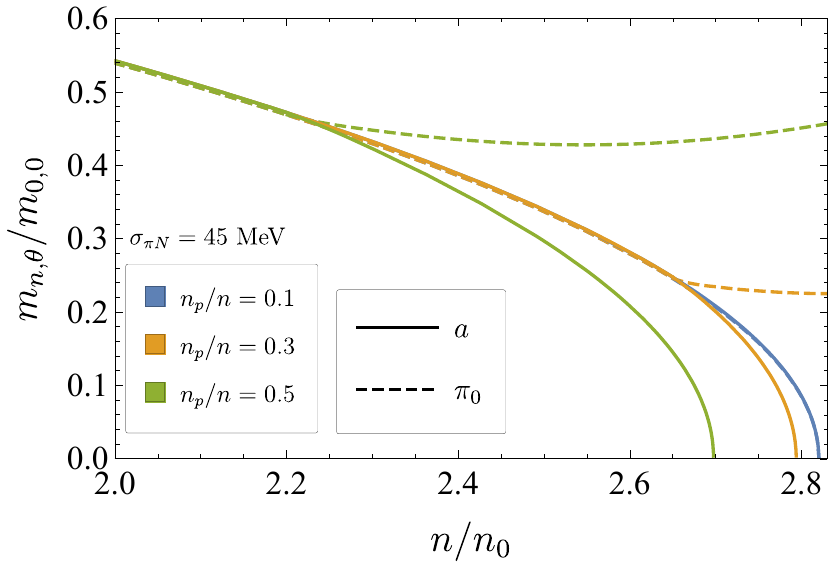}
    \caption{Numerical result for the neutral pion (dashed line) and axion (solid line) masses normalized to their $n = \theta=0$ values as a function of density $n/n_0$ for $\sigma_{\pi N} = 30 \MeV$ (left panel) and $\sigma_{\pi N} = 45 \MeV$ (right panel). The blue, orange and green curves correspond to fixed values of the proton fraction $n_p/n=0.1,\,0.3$ and $0.5$, respectively. We consider densities in the region $n < n_c \equiv n_0/b_1 \approx 2.8n_0 \, (45 \MeV/\sigma_{\pi N})$ for the corresponding values of $\sigma_{\pi N}$. The effect of kaon condensation is to eventually increase the neutral pion mass, while the axion becomes lighter and even massless at some density below $n_c$. Above this density the axion field is therefore unstable around $\vev{a}=0$ and axion condensation is expected to occur.}
    \label{kaonCondensationMasses}
\end{figure}
Note that only the neutral pion mass depends on the charge chemical potential, and that such dependence enters along with kaon condensation. The effect of a non-vanishing $\theta$ on $(m^2_{\pi_0})_{n,\theta}$ therefore depends on the relative size of $m_s/\bar{m} \approx 27$ and $2 \mu^2/(m_\pi^2\zeta_{\bar{u}u})$, which enter with opposite signs. Note in this regard that while we use the leading order result for the quark condensate ratio $\zeta_{\bar{u}u}$, we did not perform an expansion in density. This is because, as we advanced at the beginning of this section and as explicitly shown in \Fig{muandthetaPlots}, when kaon condensation sets is we have $n/n_0 > 1$. Then, when $m_s /\bar{m} > 2 \mu^2/m_\pi^2\zeta_{\bar{u}u}$, the coefficient of $\theta$ is negative and $\pi_0$ becomes lighter as kaon condensation sets in. 
Such a decrease could potentially lead to an instability and CP violation in the neutral sector.
However, in the opposite case, when $m_s /\bar{m} < 2 \mu^2/(m_\pi^2\zeta_{\bar{u}u})$, which occurs at larger densities where $\zeta_{\bar{u}u}$ is small and $\mu/m_\pi$ large (see \Fig{muandthetaPlots}), the coefficient of $\theta$ is positive and $\pi_0$ becomes heavier in the kaon-condensed phase.
As opposed to the $\pi_0$, the $\eta$ mass depends only weakly on $\theta$. 
In \Fig{kaonCondensationMasses} we plot the numerical result for $(m^2_{\pi_0})_{n,\theta}$ as a function of density for fixed values of proton fraction, using the numerical results for $\theta(n)$ and $\mu(n)$ displayed in \Fig{muandthetaPlots}. We find that the $\mu^2$ contribution leads to an increase in the mass of the neutral pion, similar to the effect of pion condensation in the simplified case discussed in~\Sec{MesonCondensation}. Finally, the axion mass is independent of $\mu$ and decreases with the size of the kaon condensate. Interestingly, the negative coefficient of $\theta^2$ is enhanced as density increases, since then $\zeta_{\bar{u}u}$ becomes smaller. As shown in \Fig{phasePlot}, this behavior eventually results in axion condensation at large densities, yet before the quark condensate vanishes. In this phase CP is thus spontaneously broken in the neutral sector.

Lastly, we note that one should be wary of the fact that for the quark condensates we included only density corrections at linear order and disregarded higher order corrections. As discussed at the end of the previous section, for densities $n \sim n_c$, these corrections are in fact important. In this regard, we would like to stress the fact that while our results might not be trustable at the quantitative level, this does not necessarily make an axion-condensed phase less likely. First, let us note that qualitatively we expect $(m^2_{a})_{n,\theta}$ to decrease with $n$ even when considering higher-order corrections to $\zeta_{\bar{u}u}(n)$. 
Second, to further support our claim, let us consider the limit of maximal kaon condensation, i.e.~$\theta \to \pi/2$, where
\begin{align}
(m^2_{a})_{n,\pi/2} \approx (m_{a}^2)_0(\zeta_{\bar{u}u}-\zeta_{\bar{s}s}) \approx -(m_{a}^2)_0\, b_1 \,\frac {n}{n_0}\,.
\end{align}
If one ignores the density dependence of the condensates by taking $b_1= 0$, this result is consistent with the naive expectation of a vanishing axion mass for $\cos\theta \to 0$, since $(m_a^2)_\theta \propto \Tr[\Sigma_0+\Sigma_0^\dagger] \propto \cos\theta$, as we showed in~\Sec{MesonCondensation}. However, as discussed in~\Sec{qqFiniteDensitySubSection}, a background of protons and neutrons makes $\zeta_{\bar{u}u} < \zeta_{\bar{s}s} \approx 1$, in other words $b_1>0$, this implies a negative axion mass for large kaon condensates, where it becomes energetically favorable for the axion field to develop a non-vanishing expectation value.

\subsection{Axion couplings} \label{axionCouplings}

Let us now turn our attention to the couplings of the axion to QCD matter at finite density. These couplings are of special importance, as they are a crucial input in the astrophysical axion bounds~\cite{Raffelt:2006cw}, in particular regarding supernovae and neutron star cooling, see e.g.~\cite{Chang:2018rso,Carenza:2019pxu,Bar:2019ifz,Potekhin:2017ufy} for recent works on the subject.

\subsubsection*{In vacuum}

These coupling have been precisely calculated including RGE effects~\cite{diCortona:2015ldu} and more recently at next-to-next-to-leading order in ChPT~\cite{Vonk:2020zfh,Lu:2020rhp}. We review here how the couplings to protons and neutrons are derived in ChPT, following closely the discussion in~\cite{diCortona:2015ldu}. 
The relevant part of the low-energy effective Lagrangian for the isospin doublet $N = (p,n)^T$ in the non-relativistic approximation is given by
\begin{align}
\mathcal{L}^{(1)}_{\pi N} \supset \bar{N} i v_\mu D^\mu N + g_A  \bar{N} S^\mu u_\mu  N +  g^{i}_0  \bar{N} S^\mu \hat{u}^i_\mu  N  + \dots \,,
\label{ChPT_nr}
\end{align}
where we omitted mass terms proportional to the axion-dependent quark mass matrix $M_a$, since there are no linear interactions coming from them if CP is preserved, see the discussion on CP violation at the end of this section. We also omitted higher-order terms in the expansion in (spatial) momenta, $k/\Lambda_{\chi} \sim k/M_B$. $v_\mu$ is the velocity of the non-relativistic fields, which satisfy $v_\mu \gamma^\mu N = N$, while $S_\mu$ is the spin operator, $ \bar{N}S_\mu N = \frac12 \bar{N}\gamma_\mu \gamma_5 N = \frac12\bar{N}(i\gamma_5 \sigma_{\mu\nu}v^\nu)N$. $D_\mu$ is the usual covariant derivative of ChPT, $D_\mu N = (\partial_\mu + e_\mu) N$ with the chiral connection $e_\mu = \frac12 (\xi^\dagger \nabla_\mu \xi + \xi \nabla_\mu \xi^\dagger)$, while the vielbein is $u_\mu = i (\xi \nabla_\mu \xi^\dagger - \xi^\dagger \nabla_\mu \xi)$, where $\xi = \exp[i \pi^a \sigma^a/2 f_\pi]$. Here we have introduced the derivative $\nabla_\mu$, which contains the external (isospin) axial and vector currents, i.e.~$\nabla_\mu \xi = \partial_\mu \xi - i (V_\mu - A_\mu) \xi$ and $\nabla_\mu \xi^\dagger = \partial_\mu \xi^\dagger - i (V_\mu + A_\mu) \xi^\dagger$.~\footnote{We have also introduced the Goldstone matrix $\xi$, because when the QCD orientation is trivial, $\xi_L = \xi_R^\dagger = \xi$. Besides, note that in our convention $u_\mu$ is twice that used in~\cite{diCortona:2015ldu}, following the standard in the ChPT literature, see e.g.~\cite{Bernard:1995dp} (however, in this reference the roles of $\xi$ and $\xi^\dagger$ are interchanged with respect to ours).} Finally, $\hat{u}^i_\mu$ is associated to the (isospin) axial scalar current, $\hat{u}^i_\mu = 2 \hat A^i_\mu$,~\footnote{In case the $\eta'$ was light, then $\hat{u}^i_\mu \propto \partial_\mu \eta'$.} where the index $i = (u + d, s)$ runs over iso-scalar quark combinations.

Because of its UV couplings to the axial currents made out of quarks, \Eq{DerCouplings}, the axion enters \Eq{ChPT_nr} as an external axial current, with components in both the iso-vector and iso-scalar directions. Therefore, with the inclusion of the axion, one finds
\begin{align}
u_\mu &= \left(\frac{\partial_\mu\pi_i}{f_\pi}\right)\sigma_i + c_-^{\IR}\left(\frac{\partial_\mu a}{f_a}\right)\sigma_3+O\left(\frac{\pi_i^2 \partial_\mu \pi_j}{f_\pi^3}\right)\,, \\
\hat{u}_\mu &= ({c}^{\IR}_{+},{c}^{\IR}_{s}) \left(\frac{\partial_\mu a}{f_a}\right)
\end{align}
with $c_\pm^{\IR} \equiv (c_u^{\IR} \pm c_d^{\IR})/2$ and where we have written $\hat{u}_\mu$ explicitly as a two-component vector. 
The couplings ${c}^{\IR}_{u,d,s}$ are related to the UV couplings by
\begin{align}
{c}^{\IR}_q = 
C_{qq'}({c}^{0}_{q'}-[Q_a]_{q'})\,.
\end{align}
where the matrix $C_{qq'}$ accounts for renormalization group evolution (RGE)~\cite{diCortona:2015ldu},
\begin{align}
C_{qq'} = 
\begin{cases}
0.975 \;\;\;\;\; &q = q'
\\
-0.024 & q\neq q'
\end{cases}\,,
\label{Cqq}
\end{align}
see~\cite{diCortona:2015ldu} for a detailed derivation.

We recall from \Sec{mu0section} that the ${c}^{0}_q$ are the UV model-dependent couplings of the axion to SM axial quark currents, while the matrix $Q_a$ was introduced in order to remove the $a G \tilde G$ term. To further eliminate all axion-pion mixing, we chose a particular rotation matrix $Q^*_a$, which at zero density (denoted by the ``0'' subscript) is given by
\begin{align}
&(Q^*_{a})_0 = \frac{\text{Diag}[1,z,zw]}{1+z+zw}\,,
\end{align}
with~\cite{Tanabashi:2018oca}
\begin{align}
&z \equiv \frac{m_u}{m_d}=0.47^{+0.06}_{-0.07}\,, \;\;\; w\equiv\frac{m_d}{m_s} = (17-22)^{-1} \,.
\end{align}
The coefficients $g_A$ and $g^i_0$ in \Eq{ChPT_nr} are given by linear combinations of hadronic matrix elements encoding the contribution of a quark $q$ to the spin operator of the proton,
\begin{align}
&g_A = \Delta u-\Delta d \,, \;\;\; g^{i}_0 =( \Delta u+\Delta d,\Delta s) \,, \;\;\; S^\mu\Delta q \equiv \vev{p | \bar{q} \gamma^\mu \gamma^5 q | p} \,.
\end{align}
The axial-vector coupling $g_A$ have been precisely measured in $\beta$-decay experiments, while $g^{ud}_0$ and $\Delta s$ have been be extracted from lattice calculations, where in both cases isospin breaking effects are neglected. Their values at zero density are~\cite{Tanabashi:2018oca,diCortona:2015ldu} 
\begin{align}
(g_A)_0 =  1.2723(23)\,, \;\;\;
(g^{ud}_0)_0 = 0.521(53)\,, \;\;\;
(\Delta s)_0 = -0.026(4)\,.
\end{align}
With this information, the couplings of the axion to nucleons,
\begin{align}
\frac{\partial_\mu a}{f_a}\left(c_p\,\bar{p}\,S^\mu\,p+c_n\,\bar{n}\,S^\mu\,n\right)
\end{align}
can be finally extracted from the effective Lagrangian in \Eq{ChPT_nr},
\begin{align}
c_p &= +g_A {c}^{\IR}_{-} +g^{ud}_0  {c}^{\IR}_{+} + \Delta s {c}^{\IR}_s \,,
\\
c_n &= -g_A  {c}^{\IR}_{-}+g^{ud}_0  {c}^{\IR}_{+} + \Delta s {c}^{\IR}_s \,.
\label{zeroDensityCouplings}
\end{align}
This leads for instance to the accurate determination of the model-independent axion couplings, i.e.~those of the KSVZ or hadronic axion~\cite{Kim:1979if,Shifman:1979if}, for which $c_q^0=0$,
\begin{align}
(c_p)_0^{\KSVZ} = -0.47(3)\,, \;\;\; (c_n)_0^{\KSVZ} = -0.02(3)\,.
\label{zeroCouplings}
\end{align}
We note in particular that the axion coupling to neutrons in vacuum is suppressed with respect to the naive $O(1)$ expectation due to an accidental cancelation between $z = m_u/m_d \approx 1/2$ and the ratio of matrix elements in vacuum $\Delta u/\Delta d = (g^{ud}_0+g_A)/(g^{ud}_0-g_A) \approx -2$,
\begin{align}
\left( \frac{c_n}{c_p} \right)_0^{\KSVZ} \propto 
\frac{1+z(\Delta u/\Delta d)_0}{(\Delta u/\Delta d)_0+z}
\approx 7.6\times10^{-2} \,,
\end{align}
neglecting RGE and other subleading effects such as $m_{u,d}/m_s$ corrections.
It is precisely this cancelation that makes $c_n$ sensitive to small variations of the parameters. For example, RGE is naively an $O(10^{-2})$ effect according to \Eq{Cqq} -- however, because of the accidental cancelation of the axion-neutron coupling in the KSVZ model, it is in fact an $O(1)$ effect, $(c_n)^{\text{\tiny no RGE}}_0/(c_n)_0 \approx 1.5$. As we will be showing in the following, finite density corrections also spoil the cancellation, leading in fact to a much larger effect.

\subsubsection*{In-medium mixing angles}

The mixing angles with the neutral pions change at finite baryon density due to the change in the values of the quark condensates, as discussed in~\Sec{qqFiniteDensitySubSection}.
The $Q^*_a$ matrix that diagonalizes such mixings becomes therefore density dependent,
\begin{align}
(Q^*_a)_n= \frac{\text{Diag}[1,z Z,z Z w W]}{1+ z Z + z Z w W} \,,
\end{align}
where we defined
\begin{align}
Z \equiv \frac{\vev{\bar{u}u}_{n}}{\vev{\bar{d}d}_{n}}\,, \;\;\; W \equiv \frac{\vev{\bar{d}d}_{n}}{\vev{\bar{s}s}_{n}}\,.
\end{align}
Using \Eq{qqs} for the quark condensates at linear order in density, we find
\begin{align}
Z = 1 - 2 \, b_2 \frac{n-2 n_p}{n_0} \,, \;\;\; W = 1-\left[b_1  - b_2 \left(1-\frac{ 2n_p}{n}\right)  - b_3 \right] \frac{n}{n_0} \, ,
\end{align}
where $b_{1,2,3}$ have been defined in terms of sigma terms in \Eq{c1c2c3}. Note that the deviation of $Z$ from unity is small, being proportional to the anomalously small coefficient $b_2$, while the effects of $W$ will be suppressed by $m_{u,d}/m_s$. 
However, similar to the RGE effects discussed above, the small density effect from $Z$ gets amplified due to the cancellation structure of $(c_n)_0^{\KSVZ}$, while no large enhancement is expected in $c_p$. Indeed, one finds 
\begin{align}
(\Delta c_n)^{Z,W\neq1}_n/(c_n)_0^{\KSVZ} & \approx 40 \% \,  \frac{n-2n_p}{n_0} \,, \\
(\Delta c_p)^{Z,W\neq1}_n/(c_p)_0^{\KSVZ} & \approx - 2.5 \% \,  \frac{n-2n_p}{n_0} \,.
\end{align}
Note that both the $O(1)$ correction to $c_n$ and the $O(10^{-2})$ correction to $c_p$ fall within the uncertainty range of the KSVZ axion couplings in vacuum given in \Eq{zeroCouplings}.

\subsubsection*{In-medium matrix elements}

The hadronic matrix elements $\Delta u$, $\Delta d$ and $\Delta s$ are also density dependent quantities in medium. In particular, the combination $g_A \equiv \Delta u-\Delta d$, which fixes the coupling of the pions to nucleons, is known to get quenched inside nucleons~\cite{Towner:1987zz}. This was observed from the reduced rates for $\beta$-decay in various large nuclei~\cite{Brown:1985zz}, suggesting that in medium $(g_A)_{n_0/2}/(g_A)_{0} \approx 0.75$, with $n_0/2$ being the typical baryon density around the surface of such large nuclei. 

The in-medium change in the effective axial coupling can be derived from the following higher-order terms in the non-relativistic baryon chiral Lagrangian~\cite{Entem:2002sf,Epelbaum:2002vt}~\footnote{Note that in the literature these terms usually appear with dimensionful coefficients $c_3$ and $c_4$. Here we normalized them to the cutoff of ChPT, $c_i \equiv \hat{c}_i \Lambda_\chi$.}
\begin{align}
\mathcal{L}^{(2)}_{\pi N} &\supset  \frac{\hat{c}_3}{\Lambda_\chi}\bar{N} u^\mu u_\mu N+\left( \frac{\hat{c}_4}{\Lambda_\chi}+\frac{1}{4M}\right)\bar{N} \, [S^\mu,S^\nu] u_\mu u_\nu \, N\,, \label{pionExchange}
\\
\mathcal{L}^{(1)}_{\pi NN} &\supset -\frac{c_D}{2 f_\pi^2 \Lambda_\chi} (\bar{N}N) (\bar{N}\,S^\mu u_\mu\, N)\,. \label{contactTerm}
\end{align}
The density dependence of $g_A$ was originally calculated in~\cite{Menendez:2011qq} and used recently to explain the apparent discrepancy in $\beta$-decay rates in large nuclei~\cite{Gysbers:2019uyb}.~\footnote{This change in $g_A$ does not only affect the axion but also neutrino dynamics in supernovae (see e.g.~\cite{Janka:2017vlw} for a review on neutrinos in supernovae), which would be interesting to explore.} It is given, at leading order in $n/\Lambda_\chi f_\pi^2$ (recall $\Lambda_\chi \sim M_B$), by
\begin{align}
\frac{(g_A)_n}{(g_A)_0} = 1 + \frac{n}{\Lambda_{\chi} f_\pi^2} \left[ \frac{c_D}{4 (g_A)_0} - \frac{I(m_\pi/k_f)}{3} \left(2\hat{c}_4-\hat{c}_3+\frac{\Lambda_{\chi}}{2M_B}\right)\right]\,,\label{gaMedium}
\end{align}
with
\begin{align}
I(x) = 1-3x^2+3x^3\tan^{-1}\left(\frac{1}{x}\right)\,.
\end{align}
We identify two types of corrections. The terms proportional to $I(m_\pi/k_f)$ arise from (the resummation of)  pion-exchange contributions originating in the operators in \Eq{pionExchange}, where $k_f = (3 \pi^2 n/2)^{1/3} \approx (270 \MeV) \, (n/n_0)^{1/3}$ and we took the limit of vanishing momentum carried by the pion (there is little variation if instead the Fermi gas average value $p_\pi^2 = 6 k_f^2/5$ is taken). The contribution proportional to $c_D$ comes instead from the contact term in \Eq{contactTerm}, as it immediately follows from the mean field result $\vev{\bar{N}N} = n$. The values of the low energy couplings $\hat{c}_3$, $\hat{c}_4$ and $c_D$ can be extracted from experiments. In this work we use the values provided in~\cite{Gysbers:2019uyb}
\begin{align}
c_D= -0.85\pm2.15\,, \;\;\; (2 \hat{c}_4-\hat{c}_3) = 9.1\pm1.4\,, 
\label{LECs}
\end{align}
for $\Lambda_\chi= 700 \MeV$. The finite density value of axial-vector coupling that follows from \Eqs{gaMedium}{LECs} is then
\begin{align}
\frac{(g_A)_{n}}{(g_A)_0} \approx 1 - (30 \pm 20) \% \, \frac{n}{n_0} \,.
\label{gAestimate}
\end{align}
Similar to $g_A$, higher-order operators in ChPT will give rise to a density dependent modification of $g_0^{ud} \equiv \Delta u+\Delta d$ and $\Delta s$,
\begin{align}
\mathcal{L}^{(2)}_{\pi N} &\supset  \frac{\hat{c}^i_3}{\Lambda_\chi}\bar{N} u^\mu \hat{u}^i_\mu N+ \frac{\hat{c}^i_4}{\Lambda_\chi}\bar{N} \, [S^\mu,S^\nu] u_\mu \hat{u}^i_\nu \, N\,, \label{pionExchangeScalar}
\\
\mathcal{L}^{(1)}_{\pi NN} &\supset -\frac{c^i_D}{2 f_\pi^2 \Lambda_\chi} (\bar{N}N) (\bar{N}\,S^\mu \hat{u}^i_\mu\, N)\,, \label{contactTermScalar}
\end{align}
from where one could derive, analogously to $g_A$, the density corrections from pion exchange or contact terms. We parametrize our ignorance about the density dependence of the axial-scalar coupling $g^{ud}_0$ by defining $\kappa$,
\begin{align}
\frac{(g^{ud}_0)_{n}}{(g^{ud}_0)_{0}} \equiv  1+ \kappa \, \frac{n}{n_0} \,,
\label{kappaDef}
\end{align}
and, in light of \Eq{gAestimate}, we will consider the two benchmarks, $\kappa = \pm 0.3$, leading to in-medium quenching, as $(g_A)_{n}$, or enhancement. Besides, while we could use a similar parametrization for $\Delta s$, its contribution to the axion couplings to nucleons is already subleading, since $(\Delta s)_0 = O(10^{-2})$, thus we will neglect it in the following. We stress that it is certainly important to properly compute the density corrections to $g_0^{ud}$ and $\Delta s$, 
a challenging task given the expected uncertainties that would be associated with the determination of the coefficients in \Eqs{pionExchangeScalar}{contactTermScalar}. 
We wish to point out however that we find no reason for the approximate relation $\Delta u/\Delta d \approx -1/z \approx -2$ to hold even if all the relevant finite density corrections are taken into account.\\

Combining both the finite density effects discussed above, let us write the density dependent axion-nucleon couplings as the obvious generlization of \Eq{zeroDensityCouplings},
\begin{align}
(c_p)_n &= +(g_A)_n ({c}^{\IR}_{-})_n +(g^{ud}_0)_n ( {c}^{\IR}_{+})_n+ (\Delta s)_n ({c}^{\IR}_s)_n\,,
\\
(c_n)_n &= -(g_A)_n ({c}^{\IR}_{-})_n +(g^{ud}_0)_n ( {c}^{\IR}_{+})_n+ (\Delta s)_n ({c}^{\IR}_s)_n\,,
\end{align}
with
\begin{align}
({c}^{\IR}_q)_n = 
C_{qq'} \left ({c}^{0}_{q'-}[(Q^*_a)_{n}]_{q'} \right)\,.
\end{align}
where we recall that, since $[(Q^*_a)_{0}]_q/[(Q^*_a)_{n}]_q \sim b_2 (n/n_0) = O(10^{-2})$ for $q = u,d$, the main effect of a baryonic background comes via $(g_A)_n$ (as well as $(g^{ud}_0)_n$), a change that affects any type of axion (something fully encoded in the coefficients $({c}^{\IR}_{\pm})_n$). 

\begin{figure}[t!]
    \centering
    \includegraphics[width=0.49\textwidth]{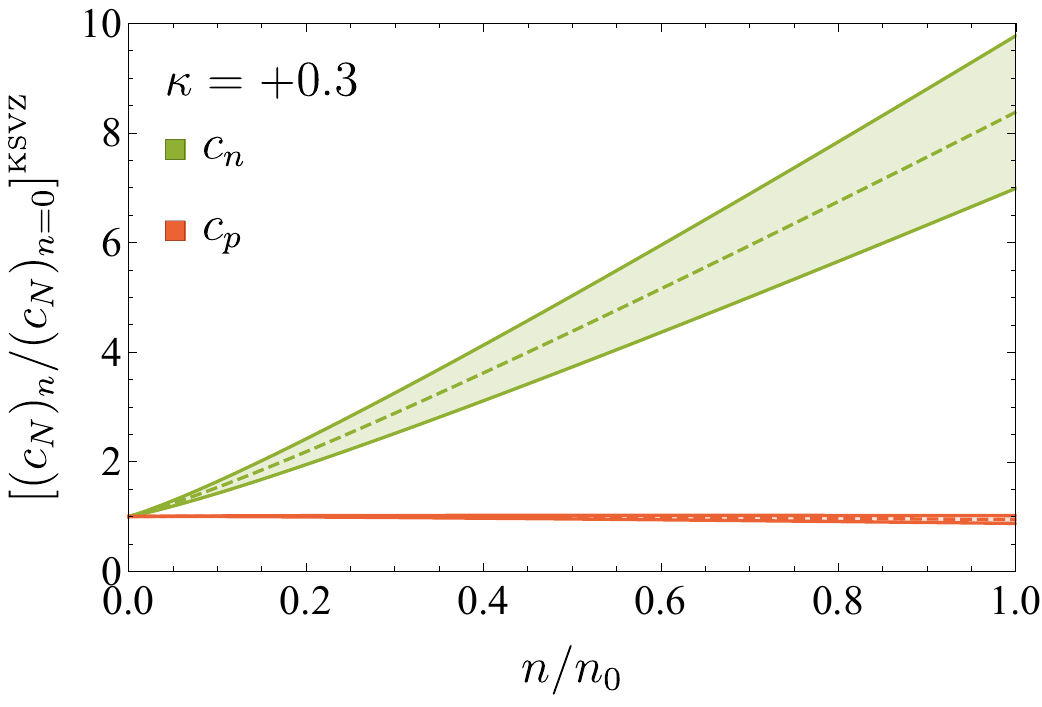}
    \hspace{0.cm}
    \includegraphics[width=0.49\textwidth]{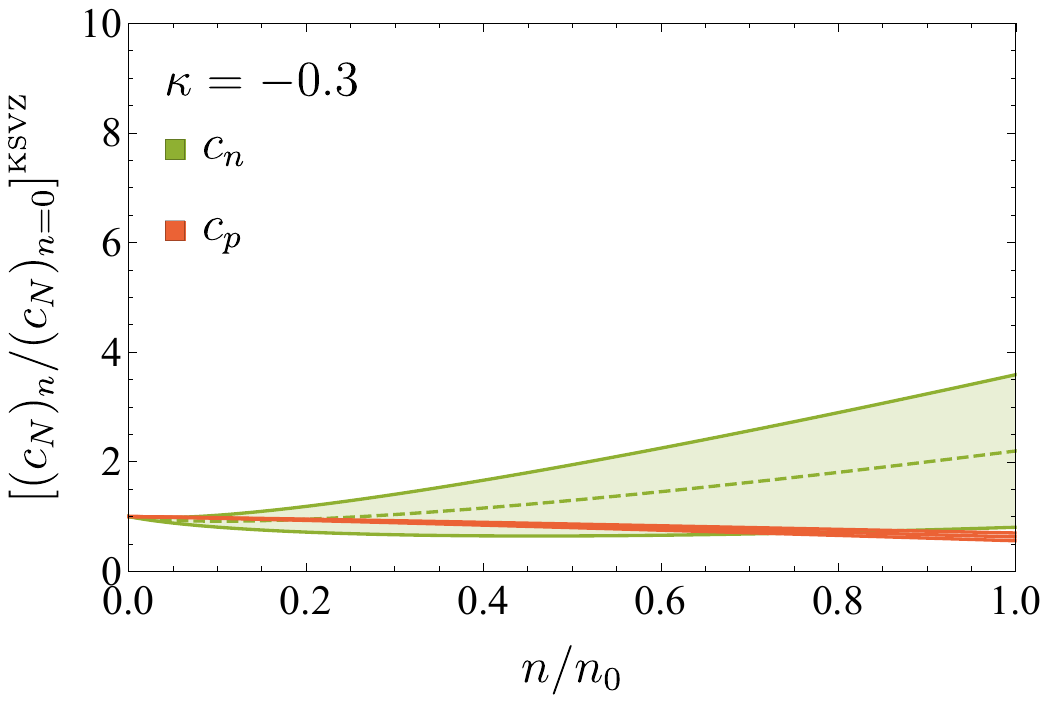}
    \caption{The model-independent axion coupling to neutrons (protons) plotted in green (red) as a function of $n/n_0$. The couplings are normalized to the zero density values, \Eq{zeroCouplings}.
    The bands represent the uncertainty in the couplings due to the coefficients in \Eq{LECs}.
    The density dependence of $g^{ud}_0$ is parameterized by $\kappa$, see \Eq{kappaDef}. We consider two benchmark cases, $\kappa=+0.3$ (left panel) and $\kappa=-0.3$ (right panel).}
    \label{withMixing}
\end{figure}

To highlight the main point of this analysis, namely that the couplings of the axion to nucleons significantly change at finite density, we plot in \Fig{withMixing} the ratio of the model-independent but density-dependent axion couplings to nucleons (including RGE), normalized to the vacuum values, as a function of $n/n_0$. As argued above, for such a hadronic axion the finite density effects are most striking, because the accidental suppression of the in-vacuum axion-neutron coupling is gone. 

In the left panel, where we take $\kappa$ positive, the $O(1)$ modification of $(g_A)_n$ and $(g^{ud}_0)_n$ translates into an $O(10)$ enhancement of $(c_n)_n^{\KSVZ}$ at nuclear-saturation densities. In case of a negative $\kappa$, the ratio 
$\Delta u/\Delta d$ remains similar to its $n=0$ value, and the accidental cancellation persists even after including in-medium effects leading only to $O(1)$ modification of $c_n$, although with large uncertainties.
For the coupling to protons we find the opposite behaviour: for $\kappa > 0$ the increase in $(g^{ud}_0)_n$ compensates for the decrease in $(g_A)_n$, such that the coupling is almost unchanged at saturation density. On the other hand, for $\kappa < 0$, both $(g^0)_n$ and $(g^{ud}_0)_n$ decrease, which leads to an $O(1)$ decrease of $c_p$. 

\subsubsection*{In kaon- and axion-condensed phases}

Let us briefly comment on the couplings of the axion when this acquires a non-trivial background value, which is the most interesting potential consequence of kaon condensation. If $\vev{a} \neq 0$ in medium, CP is violated in the neutral scalar sector and the axion acquires scalar-like couplings to nucleons, 
\begin{align}
y_{a\bar NN} \, a \bar N N,
\end{align} 
see e.g.~\cite{Moody:1984ba,Raffelt:2012sp}. While the effects of these couplings certainly deserve further investigation, in particular in the context of dense systems such as neutron stars, we only wish to point out that they are proportional to the quark masses, 
\begin{align}
y_{a \bar NN} \sim \vev{a} \, \sigma_{\pi N}/f_a^2 \sim \vev{a}\, m_{u,d}/f_a^2,
\end{align}
since in the chiral limit the expectation value of the axion is not physical.~\footnote{Axion CP-violating (self-)interactions enable the axion to mediate forces between neutron stars as investigated in~\cite{Hook:2017psm}.}

%%%%%%%%%%%%%%%%%%%%%%%%%%%%%%%%%%%%%%%%%%

\section{CFL phase} \label{CFL}

In this section we make a jump to asymptotically large baryon densities, or equivalently large quark chemical potentials, $\mu_q \gg \Lambda_\chi$ ($\mu_q \equiv \mu_u = \mu_d = \mu_s$). At such high densities, two quark around the highly energetic Fermi surface interact weakly via tree-level gluon exchange, interactions which can be effectively parametrized below the Fermi momentum by 4-Fermi operators. Such operators, when in the attractive color $\bf \bar{3}$ channel, become relevant for back-to-back scattering as one approaches the Fermi surface (see e.g.~\cite{Polchinski:1992ed,Kaplan:2005es}), leading to the formation of diquark pairs and ultimately to color superconductivity~\cite{Alford:1997zt,Rapp:1997zu,Alford:1998mk,Son:1998uk}, see also~\cite{Alford:2007xm} for a comprehensive review. Such a diquark pairing manifests itself in the form of a diquark condensate $\vev{q_L C q_L}$, which leads to the color-flavor-locked symmetry breaking pattern
\begin{align}
SU(3)_c \times SU(N_f)_L \times SU(N_f)_R \times  U(1)_{B}  \times U(1)_{A} \to SU(N_f)_{L+R+c}\,. \label{CFLsymmetry}
\end{align}
The condensates are given by~\cite{Alford:2007xm}
\begin{align}
\vev{q_L^{ia} C q_L^{jb}} = \left( \epsilon^{abc}\epsilon^{ijk} \vev{\Delta^{\bar 3}_L}_{kc} + \vev{\Delta^{6}_L}^{ij,ab} \right) \frac{3 \sqrt{2} \pi}{g_s} \frac{\mu_q^2}{2 \pi} \,,
\\
\vev{q_R^{\bar{i} a} C q_R^{\bar{j} b}} = \left( \epsilon^{abc}\epsilon^{\bar{i}\bar{j}\bar{k}} \vev{\Delta^{\bar 3}_R}_{\bar{k}c} + \vev{\Delta^{6}_R}^{\bar{i}\bar{j},ab} \right) \frac{3 \sqrt{2} \pi}{g_s} \frac{\mu_q^2}{2 \pi} \,,
\end{align}
where $i,j,k$ are $SU(3)_L$ indices, $\bar{i},\bar{j},\bar{k}$ are $SU(3)_R$ indices, and $a,b,c$ are $SU(3)_c$ indices, upper (lower) if in the (anti-)fundamental. The representations under the unbroken symmetries in (\ref{CFLsymmetry}) of the scalar field matrices above are
\begin{align}
& \Delta^{\bar{3}}_L: ({\bf \bar{3},\bar{3},1})_{+2,+2}\,, \quad  \Delta^{6}_L: ({\bf 6,6,1})_{+2,+2}\,,
\\
& \Delta^{\bar{3}}_R: ({\bf \bar{3},1,\bar{3}})_{+2,-2}\,, \quad  \Delta^{6}_R: ({\bf 6,1,6})_{+2,-2}\,,
\end{align}
while their expectation values, proportional to the gap parameters $\Delta_3$ and $\Delta_6$, are
\begin{align}
& \vev{\Delta^{\bar 3}_L}_{kc} = \delta_{kc} \Delta_3 \,, \quad \vev{\Delta^{6}_L}^{ij,ab} = (\delta^{ia}\delta^{jb}+\delta^{ja}\delta^{ib}) \Delta_6 \,, \label{firstEq}
\\
& \vev{\Delta^{\bar 3}_R}_{\bar{k}c} = -\delta_{\bar{k}c} \Delta_3\,, \quad \vev{\Delta^{6}_R}^{\bar{i}\bar{j},ab} = -(\delta^{\bar{i}a}\delta^{\bar{j}b}+\delta^{\bar{j}a}\delta^{\bar{i}b}) \Delta_6\,.
\end{align}
We parametrize the low-energy fluctuations of the ground state, i.e.~the NGBs associated with the symmetry breaking pattern (\ref{CFLsymmetry}) as
\begin{align}
\Delta^{\bar{3}}_L &= \xi_L^\dagger \vev{\Delta^{\bar{3}}_L} \exp\left[{2i\left(%-
\frac{\eta'}{f_{\eta'}}+\frac{H}{f_{H}}\right)}\right]\,,
\\
\Delta^{\bar{3}}_R &= \xi_R^\dagger \vev{\Delta^{\bar{3}}_R} \exp\left[{2i\left(-
\frac{\eta'}{f_{\eta'}}+\frac{H}{f_{H}}\right)}\right]\,,
\\
\Delta^{6}_L &= \xi^T_L \vev{\Delta^{6}_L} \xi_L \exp\left[{2i\left(%-
\frac{\eta'}{f_{\eta'}}+\frac{H}{f_{H}}\right)}\right]\,,
\\
\Delta^{6}_R &= \xi^T_R \vev{\Delta^{6}_R} \xi_R \exp\left[{2i\left(-
\frac{\eta'}{f_{\eta'}}+\frac{H}{f_{H}}\right)}\right]\,, \label{lastEq}
\end{align}
where 
\begin{align}
&\xi_L = \xi_R^\dagger = \exp\left[\frac{i\pi^a \lambda^a}{2 f_\pi} \right]\,.
\end{align}
The $\eta'$ and $H$ are the NGBs associated with the spontaneous breaking of $U(1)_A$ and $U(1)_B$, respectively. The NGBs associated with the breaking of color have been removed, since they are ``eaten'' by the gluons (unitary gauge). The rest of the NGBs, formally equivalent to those of the standard QCD chiral Lagrangian, are contained in the $\xi_{L,R}$ matrices, which are used to construct the following linearly-transforming color-neutral Goldstone matrix
\begin{align}
\Sigma &\equiv \xi_L \xi_R^\dagger : ({\bf 1,3,\bar{3}})_{0,0}\,,
\end{align}
similarly as we did in vacuum, $\mu_q = 0$, see~\Sec{mu0section}.

\subsection{Kinetic terms}

The kinetic terms in the CFL phase are given by
\begin{align}
\mathcal{L}^{\CFL}_{\text{kin.}} = \frac{f_\pi^2}{4}\eta^{\mu\nu}_\Sigma\text{Tr}[D_\mu \Sigma D_\nu \Sigma^\dagger]+\frac12\eta^{\mu\nu}_{\eta'}\partial_\mu \eta' \partial_\nu \eta'+\frac12\eta^{\mu\nu}_H \partial_\mu H \partial_\nu H\,,
\end{align}
with 
\begin{align}
\eta^{\mu\nu}_\varphi = \text{Diag}[1,-v^2_\varphi,-v^2_\varphi,-v^2_\varphi]\,,\quad \varphi = \Sigma,\eta',H\,.
\end{align}
We recall that the introduction of chemical potential breaks Lorentz symmetry down to spatial rotations, and the low-energy excitations, even if massless, propagate sub-luminally. These velocities, as well as the decay constants, can be calculated by matching the UV microscopic theory~\cite{Hong:1998tn} to the effective low-energy theory~\cite{Son:1999cm,Son:2000tu},~\footnote{As mentioned above, the gluons, with electric and magnetic masses $m_D^2 = g_s^2 f_\pi^2$ and $m_M^2 = v_\varphi^2 m_D^2$ respectively, are heavy and integrated out~\cite{Casalbuoni:1999wu}.}
\begin{align}
f^2_\pi = \frac{21-8 \ln2}{18} \frac{\mu_q^2}{2\pi^2}\,,\;\;\;
f^2_{\eta',H} = 18 \frac{\mu_q^2}{2\pi^2}\,, \;\;\; 
v^2_{\Sigma,\eta',H} = 1/3\,.
\end{align}
The $\Sigma$ field gets a dynamically induced chemical potential due to the non-vanishing quark masses~\cite{Bedaque:2001je}
\begin{align}
D_0 \Sigma = \partial_0 \Sigma + i \mu_L^{\text{eff}}\Sigma- i \Sigma\mu_R^{\text{eff}}\,, \label{effMu}
\end{align}
with
\begin{align}
\mu_L^{\text{eff}} = (\mu_R^{\text{eff}})^\dagger = \frac{M M^\dagger}{2\mu_q}\,.\label{muLmuR}
\end{align}
Note that even if we choose a basis in which the axion enters the CFL effective Lagrangian via an axion-dependent quark mass matrix $M_a$, as in \Eq{genericBasis}, it will not appear in such an effective chemical potential, since we can restrict ourselves to diagonal $Q_a$ matrices. In any case, for the analysis of the axion potential in the CFL phase, it will be more convenient to work in a basis where the axion is coupled to gluons, since a perturbative instanton expansion exists, being the gluons heavy and weakly coupled.

\subsection{Mass Terms}

Given the spurionic transformation properties of the quark mass $M$ in \Eq{massTransformation}, the leading order terms preserving the global symmetries in (\ref{CFLsymmetry}) are
\begin{align}
V_M^{\CFL} = A_1\epsilon^{ijk}\epsilon_{\bar{i}\bar{j}\bar{k}}\left([ \Delta^{\bar{3}\dagger}_L \Delta^{\bar{3}}_R]_{i}^{\,\,\,\bar{i}} M_{j}^{\,\,\,\bar{j}} M_{k}^{\,\,\,\bar{k}}+\hc\right)-\frac{A_2}{2} \left([ \Delta^{6 \dagger}_L \Delta^{6}_R]^{ij}_{\,\,\,\bar{i}\bar{j}} M_{i}^{\,\,\,\bar{i}}M_{j}^{\,\,\,\bar{j}} +\hc \right). \label{Vcfl}
\end{align}
Note that this potential respects $U(1)_A$ and it is generated perturbatively. Contrary to QCD in vacuum, the axial symmetry thus dictates that the leading order terms in the scalar potential are $O(M^2)$.
Using \EqsRange{firstEq}{lastEq} one finds~\cite{Son:1999cm}~\footnote{The first term in \Eq{Vcfl} can also be written as $-2A_1\Delta_3^2 (e^{{4i\eta'/f_{\eta'}}} \Tr[\tilde M \Sigma] + \hc )$, where $\tilde M = \det[M] M^{-1} = (\bar m m_s, \bar m m_s, \bar m^2)$.}
\begin{align}
V_M^{\CFL}  = &-A_1 \Delta_3^2 \left[e^{-{4i\eta'/f_{\eta'}}} \left( \text{Tr}[\Sigma^\dagger M]\text{Tr}[\Sigma^\dagger M]-\text{Tr}[\Sigma^\dagger M\Sigma^\dagger M] \right)+\hc \right]\nonumber
\\&+A_2 \Delta_6^2 \left[ e^{-{4i\eta'/f_{\eta'}}} \left( \text{Tr}[\Sigma^\dagger M]\text{Tr}[\Sigma^\dagger M]+\text{Tr}[\Sigma^\dagger M\Sigma^\dagger M] \right)+\hc \right]\,.
\end{align}
The coefficients can be computed by appropriately matching to the UV theory~\cite{Son:1999cm,Son:2000tu,Schafer:2001za},
\begin{align}
A_1 = 2A_2 = \frac{3}{4 \pi^2}\,. \label{ceof}
\end{align}
Importantly, these two coefficients enter with opposite signs in \Eq{Vcfl}. This is due to the fact that while the color ${\bf \bar 3}$ channel is attractive and lowers the total energy of the system, the color ${\bf 6}$ channel is repulsive and increases it. As a result, one finds that $\Delta_6 = 0 $ at the classical level. However, since $\vev{\Delta_{L,R}^6}$ does not break any additional symmetries compared to $\vev{\Delta_{L,R}^{\bar 3}}$, there is nothing preventing it from being generated at the quantum level in the presence of a non-vanishing $\Delta_3$. Indeed, a perturbative calculation yields~\cite{Schafer:1999fe}
\begin{align}
\Delta_6^2 = \alpha_s \frac{\ln^2 2}{162 \pi}\Delta_3^2\,,
\end{align}
where $\alpha_s = g_s^2/4\pi$.
$\Delta_3$ itself can be calculated using the so-called gap equation, in particular in the CFL phase with $N_f=3$~\cite{Alford:2007xm}
\begin{align}
\Delta_3 &=   512\pi^4 (2/3)^{5/2}
    e^{-\frac{\pi^2+4}{8}}
 2^{-1/3} \, \frac{\mu_q}{g_s^5} 
   \exp\left(-\frac{3\pi^2}{\sqrt{2}g_s}\right) \,.
\end{align}
The reason for considering the contribution of the condensate $\Delta_6$ to the potential, even though it is suppressed with respect to $\Delta_3$, comes from the hierarchy in the quark masses. Indeed, one finds e.g.~contributions from both condensates to the masses of the mesons, of order~\cite{Kryjevski:2004cw},
\begin{align}
\Delta_3 m_{u,d}^2 \sim \Delta_6 m^2_s\,.
\end{align}
The similarity of these two contributions, along with the fact that the coefficients of the respective operators (in \Eq{ceof}) come with opposite signs, can lead to non-trivial vacuum structures, as we review below.

Finally, we note that although at $O(M M^\dagger)$ there exist other operators which could be considered along with those in \Eq{Vcfl}, $\Tr[M \Sigma^\dagger M^\dagger \Sigma]$ and $\Tr[M \Sigma^\dagger] \Tr[M^\dagger \Sigma]$, these are not generated at the order we are interested in~\cite{Son:1999cm}.

\subsection{Non-perturbative Terms}\label{NonPerturbativeTerms}

Instantons explicitly break the $U(1)$ axial symmetry of QCD, also in the CFL phase~\cite{Schafer:2002ty}.  At leading order in the gap parameters, one finds the following term generated via a single t'-Hooft vertex
\begin{align}
V_{1\text{-inst.}}^{\CFL} = A_3 \left( [ \Delta^{\bar{3} \dagger}_L \Delta^{\bar{3}}_R]_{i}^{\,\,\,\bar{i}} [M^\dagger]^{i}_{\,\,\,\bar{i}}+\hc \right) = -A_3 \Delta_3^2 \, \text{Tr}[e^{-{4i\eta'/f_{\eta'}}}\Sigma M^\dagger+\hc]\,. \label{nonpert1}
\end{align}
The coefficient $A_3$ can be calculated reliably at large chemical potentials due to the screening of gluons for instantons of size $\rho \gtrsim 1/\mu_q \ll 1/\Lambda_{\text{\tiny QCD}}$, where $\Lambda_{\text{\tiny QCD}} \approx 250 \MeV$ is the QCD scale parameter,
\begin{align}
A_3 = c
  \left(6\pi\right)^3
  \frac{\Lambda_{\text{\tiny QCD}}^9}{  3\alpha_s^7\mu_q^{8}} \,,
  \label{a3pert}
\end{align}
with $c=0.155$~\cite{Schafer:2002ty,Alford:2007xm}. Given that the operator in \Eq{nonpert1} matches the leading term in the meson potential of the chiral Lagrangian at zero density, \Eq{VzeroDensity}, its coefficient can be mapped to the value of the standard quark condensate in the CFL phase
\begin{align}
\frac{\vev{\bar{q}q}_n^{\CFL}}{\vev{\bar{q}q}_0} = \frac{A_3 \Delta_3^2}{\vev{\bar{q}q}_0}  \sim 1 \times 10^{-5} \left(\frac{\Delta_3}{50 \MeV}\right)^2 \left( \frac{\pi}{\alpha_s}\right)^7\left( \frac{500~\text{MeV}}{\mu_q}\right)^8\left( \frac{\Lambda_{\text{\tiny QCD}}}{250 \MeV}\right)^9\,,
\end{align}
where we set the chemical potential to a value expected to be realized in the core of a NS, noting that $\alpha_s$ is to be evaluated at the scale $\mu_q$ and that $\Delta_3$ depends on both $\alpha_s$ and $\mu_q$. Due to the limited reliability of the perturbative result at such chemical potentials (see the discussion below), as well as to the strong dependence on $\Lambda_{\text{\tiny QCD}}$, it is clear that one cannot make a robust prediction regarding the value of the quark condensate at realistic densities, yet a strong suppression of $\vev{\bar{q}q}$ remains the most plausible outcome.

A higher-order operator that contributes to the mass of the $\eta'$ in the chiral limit appears at the two-instanton level,
\begin{align}
V_{2\text{-inst.}}^{\CFL} = \frac{1}{\Lambda^2} \left( \text{det}[\Delta^{\bar{3}\dagger}_L \Delta^{\bar{3}}_R]+\hc \right) = - \frac{2\Delta_3^6}{\Lambda^2} \cos \left(\frac{12 \eta'}{f_{\eta'}} \right)\,. \label{nonpert2}
\end{align}
Note this term matches the would-be leading potential for the standard $\eta'$ in vacuum, see footnote~\ref{footetap}.\\

Before moving to the discussion of the axion potential in the CFL phase, let us note that the matching procedure by which the coefficients of the effective CFL Lagrangian are extracted from the microscopic theory relies on perturbative calculations that have been found to be under control for $g_s \lesssim 0.8$~\cite{Rajagopal:2000rs}. Such a small coupling corresponds to very high quark chemical potentials, $\mu_q \gtrsim 10^8 \MeV$, five orders of magnitude higher than those expected at the cores of dense NSs, where $\mu_q \sim 500 \MeV$. Still, a quantitative but more importantly a qualitative understanding of the CFL phase and of the corresponding axion potential provides a solid ground from which to extrapolate to lower chemical potentials and thus to realistic densities. In fact, the qualitative features and basic symmetry structure of the CFL phase should hold down to $\mu \sim m_s^2/\Delta_3 \approx 180 \MeV \, (50 \MeV/\Delta_3)$~\cite{Alford:2007xm}.

\subsection{Axion potential}

In view of the previous discussion, in the following we examine the different axion potentials that arise by considering different hierarchies between the coefficients of the CFL operators.

\subsubsection*{Non-perturbative dominance}

In this case we assume that the non-perturbative contributions to the potential dominate over the mass terms, that is $V_{1\text{-inst.}}^{\CFL}, V_{2\text{-inst.}}^{\CFL} \gg V_M^{\CFL}$. Nevertheless, we still consider there exists a weak-coupling expansion, in the sense that the one-instanton contribution dominates over the two-instanton one, that is
\begin{align}
e^{-S_I} \gg e^{-2S_I} \sim e^{-S_{II}} \,,
\end{align}
with $S_I$, $S_{II}$ the action of the one- and two-instanton solutions, respectively. 
Given that the CFL operators in \Eq{Vcfl} are of order $V_M^{\CFL} \sim \bar{m} m_s \Delta_3^2$, where here and in the following we neglect $\Delta m = \frac12 (m_u-m_d)$, our hierarchy of potentials implies $\bar{m} A_3 \gg \Delta_3^4/\Lambda^2 \gg \bar{m} m_s$. In this case the potential, including the axion, reads
\begin{align}
V_{1+2\text{-inst.}}^{\CFL} = -A_3\Delta_3^2 \, \Tr[e^{i (a/f_a+
4 \eta'/f_{\eta'})} \Sigma^\dagger M +\hc] + \frac{\Delta_3^6}{\Lambda^2} (e^{i(2 a/f_a+
12 \eta'/f_{\eta'})} + \hc)\, .
\end{align}
After a field redefinition $\eta' \to \eta' - (f_{\eta'}/4f_a) a$, this is found to be the same as in the vacuum chiral Lagrangian with a light $\eta'$, which is minimized at the trivial vacuum, $\vev{\eta'} = \vev{a}=0$ in particular. The axion mass can be calculated by integrating out the mesons as we did in zero density. The details of this derivation can be found in ~\App{app1}. We find that the axion mass is suppressed with respect to its vacuum value by
\begin{align}
\frac{(m_a^2)_{\CFL}^{\text{\tiny NP}}}{(m_a^2)_0} = \frac{8 \Delta_3^6}{(m_\pi^2 f_\pi^2)_0 \Lambda^2} \sim 3 \times 10^{-3} \left(\frac{\Delta_3}{50 \MeV} \right)^6 \left( \frac{500 \MeV}{\Lambda} \right)^2 \,.
\end{align}

\subsubsection*{Perturbative dominance}

Let us now consider the hierarchy $V_M^{\CFL} \gg V_{1\text{-inst.}}^{\CFL} \gg V_{2\text{-inst.}}^{\CFL}$. One should first note that if the instanton terms are set to zero, the axion is massless, as can be immediately seen by using the basis defined by $Q_a=0$ in \Eq{genericBasis}; we use such a basis in the following. 
We write the potential in terms of the variables
\begin{align}
\vev{4\eta'/f_{\eta'} } \equiv \alpha\,, \;\;\; \vev{  a/f_a } \equiv \beta\,,
\end{align}
and use the ansatz~\cite{Kryjevski:2004cw}
\begin{align}
\vev{\Sigma } = \text{Diag}[e^{-i\varphi},e^{-i\varphi},e^{2i\varphi}] \begin{pmatrix}
1 & 0 & 0 \\
0 & \cos \theta & i \sin \theta \\
0 & i \sin \theta & \cos\theta 
\end{pmatrix}\,,
\end{align}
where the angles $\varphi$ and $\theta$ correspond to the expectation values $\vev{\eta/\sqrt{3}f_\pi}$ and $\vev{K_0/f_\pi}$, respectively. 
In this basis the meson potential is given by
\begin{subequations}
\begin{align}
&V_{\text{pert.,LO}}^{\CFL} = -\frac{f_\pi^2 m_s^4}{8\mu_q^2}\sin^2 \theta -4A_1 \Delta_3^2 \bar{m} m_s (\cos \theta +1) \cos (\alpha -\varphi ) \,, \label{locfl}
\\
&V_{\text{pert.,NLO}}^{\CFL} = -4A_1 \Delta_3^2 \bar{m}^2   \cos \theta  \cos (\alpha +2 \varphi )
+4 A_2 \Delta _6^2 m_s^2 \cos ^2\theta  \cos (\alpha -4 \varphi )\,, \label{nlocfl}
\\
&V_{1\text{-inst.}}^{\CFL} = -2A_3 \Delta _3^2 m_s     \cos \theta  \cos (\alpha +\beta +2 \varphi )\,. \label{instcfl}
\end{align}
\end{subequations}
where we separated LO terms of $O(\bar{m} m_s \Delta_3^2)$, from NLO terms of $O(\bar{m}^2 \Delta_3^2 \sim m_s^2 \Delta_6^2)$, and instanton-generated terms.~\footnote{We should note at this point that we did not include the neutral pion $\pi_0$ in our analysis because the corresponding first term in \Eq{locfl}, which destabilizes the potential at the origin of field space for $K^0$, vanishes for $\pi_0$.}
Minizing the LO potential $V_{\text{pert.,LO}}^{\CFL}$ with respect to $\varphi$ and $\theta$ yields
\begin{align}
\varphi &= \alpha \\
\cos\theta &= \text{Min}\left[1, \frac{16A_1 \Delta_3^2 \bar{m}}{m_s^3} \left( \frac{\mu_q}{f_\pi}\right)^2\right]
\label{res1}
\end{align}
The solution $\varphi=\alpha$ implies that the minimization of the NLO potential $V_{\text{pert.,NLO}}^{\CFL}$ with respect to $\alpha$ is found at
\begin{align}
\cos 3\alpha = \text{Sign}[A_1 \Delta_3^2 \bar{m}^2 - A_2 \Delta _6^2 m_s^2 \cos \theta ]\,. \label{res2}
\end{align}
\Eqs{res1}{res2} match the the results of~\cite{Kryjevski:2004cw}. Lastly, the instanton potential is minimized with respect to $\beta$ at
\begin{align}
\cos\beta = \text{Sign}\left[ \cos(3\alpha)\right] \,.
\end{align}
Therefore, one finds that the the axion is aligned with the $\eta'$, such that
\begin{align}
\vev{a/f_a}  = \begin{cases}
0 \,, & A_1 \Delta_3^2 \bar{m}^2  
> A_2 \Delta _6^2 m_s^2 \cos \theta \\
\pi \,, & A_1 \Delta_3^2 \bar{m}^2  
< A_2 \Delta _6^2 m_s^2 \cos \theta
\end{cases}\,,
\end{align}
while the axion mass, neglecting mixing with $\eta'$ and normalized to its vacuum value, is given by
\begin{align}
\frac{(m_a^2)_{\CFL} }{(m_a^2)_{0} } =  \frac{8 A_3 \Delta _3^2 m_s \cos \theta}{(m_\pi^2 f_\pi^2)_0 } \sim 7 \times 10^{-4} \left(\frac{\Delta_3}{50 \MeV} \right)^2 \left( \frac{A_3}{4 \times 10^{-4}\MeV}\right) \left( \frac{\cos \theta}{1}\right)
\,. \label{cflMass}
\end{align}
where we evaluated $A_3$ in \Eq{a3pert} at $\mu_q = 1 \GeV$, $\Lambda_{\text{\tiny QCD}} = 250 \MeV$ and $\alpha_s = \pi$.
We therefore find that the axion can develop a non-vanishing expectation value in the CFL phase also, when the kaon condensate is large. Up to uncertainties associated with the value of $A_3$, the axion is  significantly lighter than in vacuum.

%%%%%%%%%%%%%%%%%%%%%%%%%%%%%%%%%%%%%%%%%%

\section{Axion sourcing observables} \label{observablesss}

We briefly discuss in this section the potentially observable consequences of a non-vanishing axion condensate in NSs, where the largest baryonic densities among the stars are found. We defer to future work a more in-depth analysis of the corresponding phenomenology~\cite{vacNS}, as well as the study of the implications of the change in the axion-nucleon couplings with density, the latter particularly relevant for supernovae and NS cooling.

\begin{figure}[h!]
  \centering
    \includegraphics[width=0.7\textwidth]{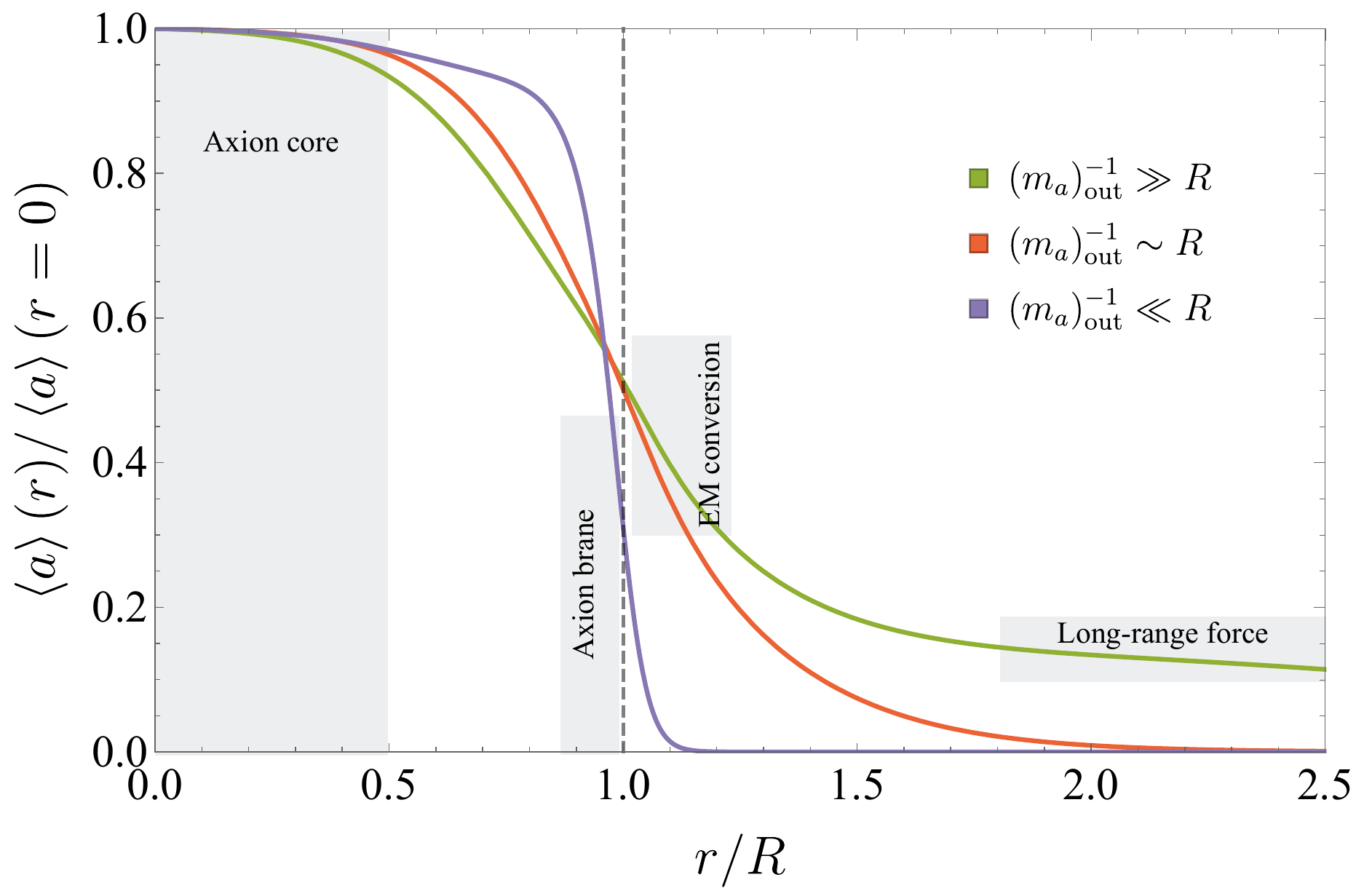}
    \caption{Sketch of the typical field configurations of a sourced axion field, see the discussion in the main text.}
    \label{obs}
\end{figure}

For simplicity, let us consider the following toy model, namely a stepwise radius-dependent axion potential
\begin{align}
V(a,r) &= \begin{cases}
f_a^2 (m_a^2)_{\text{in}} \left[\cos(a/f_a)-1\right] \;\;\;\;\; &r<r_c 
\\
-f_a^2 (m_a^2)_{\text{out}} \left[\cos(a/f_a)-1\right] &r>r_c 
\end{cases}\label{simpPot}\,, \quad f_a^2 (m_a^2)_{\text{out}} \sim m_\pi^2 f_\pi^2 \,,
\end{align}
where $m_\pi$ and $f_\pi$ are the vacuum values and we have fixed the constants such that in the decoupling limit $f_a\to \infty$, the potential vanishes. The potential grossly captures the effect of matter on the axion potential, i.e.~at a critical radius $r_c$, which is of the order of the NS radius $R$, the axion field gets destabilized and the minimum of the potential is located at $\vev{a/f_a} = \pi$. The field equation can be solved numerically and one finds the intuitive result based on energy conservation, i.e.~for the axion to get sourced the gain in potential energy needs to be enough to compensate for the gradient energy that comes with the change in field value, $\Delta V \sim (\Delta a/R)^2$, which occurs when the object is large enough compared to the de Broglie wavelength of the field inside the object~\cite{Hook:2017psm}, namely
\begin{align}
(m_a)^{-1}_{\text{in}}\lesssim r_c \sim R \,.
\label{axionsourced}
\end{align}
Let us assume this is the case for the rest of the discussion, keeping in mind that the axion mass decreases with baryon density and that in vacuum $(m_a)_{\text{out}}^{-1} \sim 16 \, {\text{km}} \, (f_a/10^{18} \GeV)$. The typical field configuration of the sourced axion is roughly 
\begin{align}
\frac{a(r)}{\pi f_a} = \begin{cases}
1 \, , &r<r_c
\\
\frac{r_c}{r}e^{-(m_a)_{\text{out}}(r-r_c)} \,, & r>r_c
\end{cases}\,.
\label{aprofile}
\end{align}
In \Fig{obs} we depict the typical field configurations of the axion sourcing, highlighting in grey the possible observable implication, to be discussed in turn below.

\subsection{Free (vacuum) energy}

The first potentially observable implication is associated with the shift in potential energy density inside the NS, 
\begin{align}
\Delta V \sim -2f_a^2 (m_a^2)_{\text{in}},
\end{align}
as a result of the axion sourcing. This effect is independent of the field configuration outside the core of the NS, namely it is independent of $\left(m_a^2\right)_{\text{out}}$. Such an energy density shift can be of considerable size compared to the energy density inside a NS, $\rho_0 \approx m_n n_0 \approx (190 \MeV)^4$. Indeed, if the axion is sourced at relatively low baryon densities, as in the kaon condensed phase (\Sec{KaonCondensation}), one expects $\Delta V \sim m_\pi^2 f_\pi^2$, which is indeed not significantly below $\rho_0$ or the energy change due to kaon condensation, of $O(m_{K}^2 f_\pi^2)$. Instead, if axion sourcing happens in the CFL phase (\Sec{CFL}), this effect is expected to be suppressed by a few orders of magnitude, see \Eq{cflMass}, and therefore likely negligible. 

A NS with a core of ``vacuum energy'' was considered as a generic scenario in \cite{Bellazzini:2015wva,Csaki:2018fls}, in the context of exotic QCD phases. 
It was found that the energy shift inside the NS leads to a significant change in the mass-radius relation of NSs, as well as to changes of the so-called chirp mass, one of main the gravitational wave observables of compact binary mergers. One could then, in a similar fashion, consider axion condensation as a possible source for such a potential energy shift.

\subsection{Axion brane}

Another interesting implication of axion condensation concerns the generation of a brane of energy density inside the NS. In particular, if the condition
\begin{align}
(m_a)^{-1}_{\text{out}} \ll R\,,
\end{align}
is met, the exponentially suppression of the axion field outside the NS is very rapid, a change that contributes to the energy density of the system in the form of a gradient energy
\begin{align}
(\vec\nabla a)^2 \sim \left(\frac{\Delta a}{\Delta r}\right)^2 \sim f_a^2(m_a)^{2}_{\text{out}} \sim m_\pi^2 f_\pi^2 \,.
\end{align}
One can think of such an abrupt change in field values as a localized (spherical) brane of energy density. The effect of such a brane has in fact not been considered in previous studies of NS structure.
Such a brane would appear as an effective discontinuity in the temporal and radial components of the metric, which because of Einstein's field equations (known as the Tolman-Oppenheimer-Volkoff equations), imply a discontinuity in the pressure and enclosed mass of the star. 

\subsection{Axion-EM conversion}

Next we consider the interplay between the EM fields of rotating NSs (i.e.~pulsars), which are the strongest found in the Universe, and the axion, in particular when 
\begin{align}
(m_a)^{-1}_{\text{out}} \gtrsim R\,,
\end{align}
such that the sourced axion field is still non-negligible in the close surroundings of the NS. 

The axion and the classical EM fields form a coupled system, as seen from the generalized form of the Maxwell equations
\begin{subequations}
\begin{align}
\vec{\nabla} \cdot {\bf{E}} &= g_{a\gamma\gamma}\, {\bf{B}} \cdot (\vec{\nabla} a)\,,
\\
\vec{\nabla} \times {\bf{B}} &=  \frac{\partial {\bf E}}{\partial t}+g_{a\gamma\gamma} \left[ {\bf{E}}\times \vec{\nabla}a-{\bf B} \dot{a}\right]\,,
\\
\Box a &=  g_{a\gamma \gamma} ({\bf{E}}\cdot {\bf{B}}) - \frac{\partial V}{\partial a} = g_{a\gamma \gamma} ({\bf{E}}\cdot {\bf{B}})-(m_a^2)_{\text{out}} a + O(a^2) \,,
\end{align}
\end{subequations}
where the last line is the axion equation of motion. The interplay between the axion and the EM field of pulsars has been actively investigated before, see e.g.~\cite{Garbrecht:2018akc,Fortin:2018ehg,Day:2019bbh,Buschmann:2019pfp} for recent works on the subject, although the effects we consider here, associated with a large classical axion field configuration also sourced by the NS, are novel.
Assuming the conventional rotating dipole model, one finds that at the surface of the NS
\begin{align}
&B_{\text{\tiny dipole}}(R) \sim B_* \sim 10^{14}\,\text{G} \sim \MeV^2\,,
\\
&E_{\text{\tiny dipole}}(R) \sim R \Omega B_* \sim 10^{-3} \left(\frac{R}{10\,\text{km}}\right)\left(\frac{\Omega}{100\,\text{Hz}}\right)B_*\,,
\end{align}
with $\Omega$ the angular velocity of the NS. Even with such large EM fields, we may still neglect the effects of the axion-photon coupling on the axion dynamics, since
\begin{align}
\frac{ g_{a\gamma \gamma} ({\bf{E}}\cdot {\bf{B}}) }{(m_a^2)_{\text{out}} \vev{a(R)}} 
\sim \frac{\alpha_{\EM} R \Omega B_*^2}{m_\pi^2 f_\pi^2} \sim 10^{-13} \left(\frac{R}{10\,\text{km}}\right)\left(\frac{\Omega}{100\,\text{Hz}}\right) \left(\frac{B_*}{10^{14}\,\text{G}}\right)^2 \,,
\label{EMneglect}
\end{align}
where we used $(m_a^2)_{\text{out}} \sim m_\pi^2 (f_\pi/f_a)^2$, $\vev{a(R)} \sim f_a$ and $g_{a\gamma \gamma} \sim \alpha_{\EM}/f_a$.
While we can safely assume that the back-reaction of the EM fields on the axion is negligible, it is also important to note that the value of $\vev{a(R)}$ decreases exponentially outside the NS, see \Eq{aprofile}, such that one could well imagine a situation where the effect of the $g_{a\gamma \gamma} ({\bf{E}}\cdot {\bf{B}})$ term is in fact comparable to the axion mass term. In this case the back-reaction of the EM fields would have to be taken into account, which is beyond the scope of this work. 

We thus treat the axion field as a rigid source of additional EM fields, which can be simply estimated as
\begin{align}
&\Delta E \sim g_{a\gamma\gamma} B_* \vev{a(R)} \sim \alpha_{\EM} B_*\,,
\\
&\Delta B  \sim g_{a\gamma\gamma} R \Omega B_* \vev{a(R)} \sim \alpha_{\EM} R \Omega B_*\,.
\end{align}
While the magnetic field receives a small correction $\Delta B/B \sim \alpha_{\EM} R \Omega \ll 1$, for the electric field
\begin{align}
\frac{\Delta E}{E} \sim \frac{\alpha_{\EM}}{R \Omega} \sim 2 \left(\frac{10\,\text{km}}{R}\right)\left(\frac{100\,\text{Hz}}{\Omega}\right) \,,
\end{align}
thus leading to an $O(1)$ enhancement around the surface of the NS. 
We note that since this correction is large, one could be concerned about whether the system can be treated perturbatively. This is in fact the case, since the higher order terms scale like
\begin{align}
E \sim R \Omega B_* (1+ \alpha^2_{\EM} + \dots ) + \alpha_{\EM} B_* (1+ \alpha^2_{\EM} + \dots )\,.
\end{align}
This means that, apart from the leading $O(\alpha_{\EM} B_*)$ correction, further contributions are subleading.

An additional sensitive observable is the dipole radiation output $P$ that is responsible for the spin-down of rotating NSs. In this case we find that $\Delta P/P \sim \alpha^2_{\EM} \ll 1$, namely there is no appreciable addition to the radiated energy due to the axion field.

\subsection{Long-range force}

Lastly, we can consider the case 
\begin{align}
(m_a)^{-1}_{\text{out}}\gg R\,,
\end{align}
even though we note that from our previous analysis of the QCD axion at finite density, we expect this regime not to be realized since $(m_a)^{-1}_{\text{out}} \lesssim (m_a)^{-1}_{\text{in}} \lesssim R$, where the last condition follows from the requirement of the axion being actually sourced, \Eq{axionsourced}.
Therefore we expect the hierarchy $(m_a)^{-1}_{\text{in}} \lesssim R \ll (m_a)^{-1}_{\text{out}}$ to arise only in non-standard scenarios, such as the one considered in~\cite{Hook:2017psm}. If that is indeed the case, the long tails of the axion field configuration lead to a long range force between the NSs, generated by the Yukawa-like potential 
\begin{align}
V \sim \frac{Q_{\text{eff}}}{r}e^{-(m_a)_{\text{out}} r}\,,
\end{align}
where $Q_{\text{eff}} = 4 \pi f_a R$ plays the role of the effective charge. This could lead to a deformation of the merger wave-form predicted by general relativity in case of NS with opposite-sign charges. A more dramatic effect would be found in the case of a repulsive force from same-sign charges, since in this case at some critical distance the axion force would dominate gravity, which could lead to halt in the merger process~\cite{Hook:2017psm}. The presence of the axion field can also lead to an additional mechanism of energy loss in NS mergers, in the form of the scalar equivalent of Larmor radiation~\cite{Huang:2018pbu}. 

%%%%%%%%%%%%%%%%%%%%%%%%%%%%%%%%%%%%%%%%%%

\section{Conclusions and Outlook} \label{conc}

In this paper we have shown how the properties of the QCD axion change with baryon chemical potential. 
Based on the non-relativistic baryon chiral Lagrangian, reliable up to (slightly above) nuclear saturation densities, we have found that the axion gets lighter in medium, albeit the reduction is below an order of magnitude. In contrast, we have found that density corrections have a large impact on the axion couplings to nucleons, in particular the model-independent (a.k.a.~KSVZ) coupling to neutrons is enhanced by up to one order of magnitude, thus becoming, in a baryonic background at nuclear saturation density, of the same order as the coupling to protons.

We have also derived the conditions under which the axion acquires a non-trivial average value as a result of kaon condensation, an hypothetical yet motivated scenario for nuclear matter at densities above nuclear saturation.

Since reliable predictions for QCD matter at baryon densities as those found in the cores of neutron stars are currently unavailable, in order to delimitate the properties of the QCD axion not only from low but from high densities as well, we have analyzed the axion potential in the color-flavor-locked phase of QCD. We have again derived the conditions under which axion condensation takes place, and found that the axion mass is generically several orders of magnitude smaller than in vacuum.

The main goal of this paper was to lay the groundwork for future phenomenological investigations of axion physics in dense systems. This is the reason why we only brushed over some of the corresponding experimental implications, focussing in particular on the sourcing of the axion. In this context, we are currently studying in detail the implications of such an axion condensate contributing to the mass of neutron stars, as well as the idea of a brane separating the inner axion core from the rest of the neutron star \cite{vacNS}, which are interesting idealizations of the richer dynamics of axions within these astrophysical objects. Another very compelling avenue concerns the interplay of the axion with the large electromagnetic fields generated by pulsars. Importantly, these are only a subset of the range of exciting consequences our analysis has uncovered. In particular, the significance of the finite density corrections on the axion couplings calls for a proper assessment of how all such type of effects (as well as finite temperature) affect the predictions of the cooling rates of supernovae and proto-neutron stars, of special significance  for the QCD axion. 

Moreover, we believe our work will be of relevance for the study of other well-motivated new physics scenarios at finite density. This is certainly the case for those realizations of the so-called relaxion that rely on QCD dynamics -- on the dependence of the quark masses on the electroweak scale -- to generate a dynamical landscape for the Higgs fields. The investigation of this and similar density-dependent landscapes and its interplay with neutron stars is something that we leave for a future publication~\cite{Balkin:2021wea}.

%%%%%%%%%%%%%%%%%%%%%%%%%%%%%%%%%%%%%%%%%%

\section*{Acknowledgments}

We would like to thank M.~Alford, L.~Fabbietti, N.~Kaiser, D.B.~Kaplan, J.I.~McDonald, G.~Raffelt, P.~Panci, G.~Villadoro and S.~Stelzl for helpful and interesting discussions. We also thank the Munich Institute for Astro- and Particle Physics (MIAPP) for hospitality in the final stages of this work. The work of RB, JS, KS and AW has been partially supported by the DFG Cluster of Excellence 2094 ORIGINS, by the Collaborative Research Center SFB1258, and by the DFG project number 378905682.

%%%%%%%%%%%%%%%%%%%%%%%%%%%%%%%%%%%%%%%%%%
%%%%%%%%%%%%%%%%%%%%%%%%%%%%%%%%%%%%%%%%%%
\newpage
\appendix

\section{Axion mass calculation with instantons} \label{app1}

We calculate the axion mass by integrating out the neutral pions $\{\pi_0,\eta, \eta' \}$ in the $N_f = 3$ chiral Lagrangian, from the potential
\begin{align}
{V}_{0} = b (\text{Tr}[\Sigma^\dagger M]+\hc) - c (e^{-i(a/f_a+\eta'/f_{\eta'})} + \hc) \,,
\end{align}
where in vacuum $b$ is given in \Eq{bchiral} and effectively $c \to \infty$, while $b = - A_3 \Delta_3^2$ and $c = \Delta_3^6/\Lambda^2$ in the CFL phase when instantons dominate. Note that the $\eta'$ is normalized  differently in the CFL phase.
This procedure produces the correct leading order result for the axion mass but neglects (some of) the subleading corrections. In the generic basis of \Eq{genericBasis}, the potential then reads
\begin{align}
V = &  2b m_u \cos \left(\frac{ Q_u a}{f_a}-\frac{\frac{\eta }{\sqrt{3}}+\pi_0}{f_{\pi }}-\frac{\eta '}{3 f_{\eta '}}\right)
+ 2b m_d \cos \left(\frac{ Q_d a }{f_a}+\frac{\pi_0-\frac{\eta }{\sqrt{3}}}{f_{\pi }}-\frac{\eta '}{3 f_{\eta '}}\right) \nonumber \\
&+ 2b m_s \cos \left(\frac{ Q_s a}{f_a}+\frac{2 \eta }{\sqrt{3} f_{\pi }}-\frac{\eta '}{3 f_{\eta '}}\right)
- 2c \cos \left( \frac{\eta'}{f_{\eta'}}-\frac{ \left(\text{Tr}[Q_a]-1\right)a}{f_a} \right)\,,
\end{align}
with $Q_a = \text{Diag}[Q_u,Q_d,Q_s]$.
We integrate $\pi_0$ out by using its equation of motion at linear order in the fields
\begin{align}
\pi_0 = \frac{  \left(m_u Q_u-m_d Q_d\right)}{ \left(m_d+m_u\right) } \left(\frac{f_\pi }{f_a}a \right)
+
\frac{ \left(m_d-m_u\right) }{3 \left(m_d+m_u\right) }\left(\sqrt{3} \eta  +\frac{f_{\pi }}{f_{\eta '}} \eta '\right) \,. 
\end{align}
Next we similarly integrate $\eta$ out
\begin{align}
\eta = &\frac{\sqrt{3}  \left(-m_d m_s Q_s+m_d m_u \left(Q_d+Q_u\right)-m_s m_u Q_s\right)}{2  \left( m_d m_s+m_um_d +m_s m_u\right)} \left(\frac{f_\pi }{f_a}a \right) \nonumber\\
&+
\frac{ \left(m_d \left(m_s-2 m_u\right)+m_s m_u\right)}{2 \sqrt{3}  \left( m_d m_s+m_u m_d+m_s m_u\right)}\left(\frac{f_{\pi }}{f_{\eta '}} \eta '\right)\,.
\end{align}
Finally, we integrate out $\eta'$
\begin{align}
\eta' = \frac{- b m_d m_s m_u \text{Tr}[Q_a]+c \left( m_d m_s+m_u m_d +m_s m_u\right) \left(\text{Tr}[Q_a]-1\right)}{ - b m_d m_s m_u+c \left( m_d m_s+m_u m_d +m_s m_u\right)}\left(\frac{f_{\eta'} }{f_a}a \right)\,.
\end{align}
The potential is minimized around $\vev{a} = \vev{\pi_0} =\vev{\eta} =\vev{\eta'}=0$ and we find the following axion mass
\begin{align}
m_a^2 &=  \frac{ - 2 b c\, m_u  m_d}{f_a^2 (m_u+m_d)\left(c\left[1+  \frac{m_u m_d}{m_s(m_u+m_d)}\right] - b \left[\frac{m_d  m_u}{m_u+m_d}\right] \right)}\,.
\label{lowDensMass}
\end{align}

Once we have diagonalized the mass matrix, one could be concerned with the effect of the $O(f_\pi/f_a)$ kinetic mixing which is generically induced by the derivative couplings of the axion. Let us explictly show that this does not affect the leading order results for the axion mass. Our starting point, without loss of generality, is the following Lagrangian
\begin{align}
\mathcal{L} = 
\frac12 m_\pi^2 \begin{pmatrix}
\vec{\pi}& a
\end{pmatrix}
\begin{pmatrix}
b &0 \\
0 & c \xi^2
\end{pmatrix}
\begin{pmatrix}
\vec{\pi}\\
a
\end{pmatrix}
+
\frac12  \begin{pmatrix}
\partial_\mu \vec{\pi}& \partial_\mu  a
\end{pmatrix}
\begin{pmatrix}
1 &\xi \vec{d} \\
\xi \vec{d}^T & 1
\end{pmatrix}
\begin{pmatrix}
\partial^\mu  \vec{\pi}\\
\partial^\mu  a
\end{pmatrix}\,, \label{originalForm}
\end{align}
where $b = \text{Diag}[b_1,..,b_n]$ is an $n\times n$ diagonal matrix of $O(1)$ numbers, $\vec{d}$ is a vector of $n$ $O(1)$ numbers, $c$ is an $O(1)$ number, and $\xi \equiv f_\pi/f_a$ is our expansion parameter. Let start by performing the orthogonal rotation $R_1$ in the meson subspace, such that
\begin{align}
R_1 \vec{d} = (0, \dots ,|d|).
\end{align}
We rewrite the Lagrangian in this basis
\begin{align}
\mathcal{L} = 
\frac12 m_\pi^2 \begin{pmatrix}
\vec{\pi}& a
\end{pmatrix}
\begin{pmatrix}
R_1bR_1^T &0 \\
0 & c \xi^2
\end{pmatrix}
\begin{pmatrix}
\vec{\pi}\\
a
\end{pmatrix}
+
\frac12  \begin{pmatrix}
\partial_\mu \pi_1&...&\partial_\mu \pi_n& \partial_\mu  a
\end{pmatrix}
\begin{pmatrix}
1 &&& \\
 &\ddots&& \\
&&1&|d|\xi \\
 &&|d|\xi&1 \\
\end{pmatrix}
\begin{pmatrix}
\partial^\mu \pi_1\\
\vdots \\
\partial^\mu \pi_n\\
 \partial^\mu  a
\end{pmatrix}\,.
\end{align}
We diagonalize and canonically normalize the lower $2 \times 2$ block in the second term by rotating and rescaling the fields
\begin{align}
\begin{pmatrix}
\pi_n\\
a
\end{pmatrix}
=
\underbrace{\frac{1}{\sqrt{2}} \begin{pmatrix}
1 & 1 \\
-1 & 1
\end{pmatrix}}_{\equiv R_2}
\underbrace{\begin{pmatrix}
\frac{1}{\sqrt{1-|d |\xi}} & 0 \\
0  & \frac{1}{\sqrt{1+|d | \xi }}
\end{pmatrix}}_{\equiv T}
\begin{pmatrix}
\bar\pi_n\\
\bar a
\end{pmatrix} \,.
\end{align}
Our mass matrix in the new basis now reads
\begin{align}
m _\pi^2 T R_2^T \begin{pmatrix}
[R_1bR_1^T]_{nn} &0 \\
0 & c \xi^2
\end{pmatrix} R_2 T\,.
\end{align}
$T$ can be expanded $T = \mathbb{1}+ \frac12 |d| \xi \sigma_3 + O(\xi^2)$. At leading order $T=\mathbb{1}$ and the mass matrix can be re-diagonalized by performing the inverse orthogonal rotation $R_2^{-1}$, bringing it back to the diagonal form of \Eq{originalForm}. One concludes that the axion mass does not receive any leading order correction due to the kinetic mixing. 

\section{Baryon-ChPT with non-trivial vacuum alignment} \label{appChPT}

We generalize the $N_f=3$ chiral Lagrangian with baryons for a non-trivial ground state orientation, $\Sigma_0 \neq 1$ with $\Sigma_0^\dagger \Sigma_0 = \mathbb{1}_3$, e.g.~in the kaon-condensed phase
\begin{align}
\Sigma_0(\theta) = \left(
\begin{array}{ccc}
\cos \theta  & 0 & i \sin \theta  \\
0 & 1 & 0 \\
i \sin \theta  & 0 & \cos \theta  \\
\end{array}
\right)\,.
\end{align}
We denote $\Sigma_0(\theta/2) \equiv  \xi_0(\theta)$ (such that $\xi_0^2 = \Sigma_0$) and drop for brevity the explicit $\theta$-dependence.
The standard $SU(3)_L \times SU(3)_R$ generators are given by the Gell-Mann matrices
\begin{align}
T_L^a = \lambda^a\,, \;\;\; T_R^a =\lambda^a \,.
\end{align}
It should be understood that the $L$ and $R$ operators act on different indices and therefore commute. 
We define the following rotated generators
\begin{align}
(T_L^a)_\theta = \xi_0 (T_L^a) \xi_0^\dagger\,, \;\; (T_R^a)_\theta =\xi_0^\dagger (T_R^a) \xi_0\,,
\end{align}
The broken and unbroken generators are given by
\begin{align}
X^a = (T_L^a)_\theta-(T_R^a)_\theta\,, \;\;\; T^a = (T_L^a)_\theta+(T_R^a)_\theta\,,
\end{align}
respectively.
The fluctuation around the vacuum are parametrized by the Goldstone matrices
\begin{align}
&\xi_L = e^{i \frac{\pi^a}{2f_\pi} (T_L^a)_\theta} = \xi_0 \exp\left[\frac{i\pi^a \lambda^a}{2f_\pi} \right] \xi_0^\dagger\, , \quad
\xi_R = e^{-i \frac{\pi^a}{2f_\pi} (T_R^a)_\theta} = \xi_0^\dagger \exp\left[-\frac{i\pi^a \lambda^a}{2f_\pi} \right] \xi_0\,,
\end{align}
with transformation properties 
\begin{align}
\xi_L \to L \xi_L V_\theta^\dagger \,, \;\;\; \xi_R = R \xi_R V_\theta^\dagger \,,
\end{align}
with $V_\theta$ a NGB-dependent transformation under the unbroken $SU(3)$ subgroup of $SU(3)_L \times SU(3)_R$, the transformations under the latter denoted by $L$ and $R$ respectively.
As usual, it is convenient to construct
\begin{align}
\Sigma = \xi_L \Sigma_0 \xi_R^\dagger = \xi_0 \exp\left[\frac{i\pi^a \lambda^a}{f_\pi} \right] \xi_0 \,,
\end{align}
which transforms as $\Sigma \to L \Sigma R^\dagger$. Following standard notation,
\begin{align}
\pi^a \lambda^a = \sqrt{2} \left(
\begin{array}{ccc}
\frac{\pi_0}{\sqrt{2}} + \frac{\eta}{\sqrt{6}} & \pi^+ & K^+ \\
\pi^- & -\frac{\pi_0}{\sqrt{2}} + \frac{\eta}{\sqrt{6}} & K_0 \\
K^- & \bar{K}_0 & -\sqrt{\frac{2}{3}} \eta  \\
\end{array}
\right)\,.
\end{align}
We introduce the ($\theta$-rotated) baryon octet as linearly-transforming fields, $\hat B_{L,R}^\theta$,
\begin{align}
\hat B^\theta_L \to L \hat B^\theta_L L^\dagger \,, \;\;\; \hat B^\theta_R = R \hat B^\theta_R R^\dagger \,,
\end{align}
where we use the $\theta$-superscript because the finite-density backgrounds we consider consist of a non-vanishing ensemble of the standard (non-rotated) baryons, given by
\begin{align}
\hat B_L = \xi_0^\dagger \hat B^\theta_L \xi_0 \,, \;\;\; \hat B_R = \xi_0 \hat B^\theta_R \xi_0^\dagger \,,
\label{bRotate}
\end{align}
with the usual parameterization 
\begin{align}
B_{L,R} = \left(
\begin{array}{ccc}
\frac{\Sigma_0}{\sqrt{2}} + \frac{\Lambda}{\sqrt{6}} & \Sigma^+ & p \\
\Sigma^- & -\frac{\Sigma_0}{\sqrt{2}} + \frac{\Lambda}{\sqrt{6}} & n \\
 \Xi^- & \Xi_0 & -\sqrt{\frac{2}{3}} \Lambda  \\
\end{array}
\right)_{L,R} \,.
\end{align}
The Lagrangian in this basis is given by
\begin{align}
\mathcal{L} &=\mathcal{L}^0_{\Sigma}+\mathcal{L}^0_{B}+\mathcal{L}_M+\mathcal{L}^0_{ \ell}\,,
\label{linearL}
\\
\mathcal{L}^0_{\Sigma}&= \frac{f_\pi^2}{4}\Tr[\partial_\mu \Sigma^\dagger \partial^\mu \Sigma]\,,
\\
\mathcal{L}^0_B &= i \Tr[\bar{\hat B}^\theta_L \gamma^\mu \partial_\mu \hat B^\theta_L] + i \Tr[\bar{\hat B}^\theta_R \gamma^\mu \partial_\mu \hat B^\theta_R] - M_B \Tr[\bar{\hat B}^\theta_L \Sigma \hat{B}^\theta_R \Sigma^\dagger+\hc] \,,
\\
\mathcal{L}_M =& -\frac{\vev{\bar{q}q}_0}{2} \Tr[\Sigma^\dagger M] \nn
\\
&+ a_1 \Tr[\bar{\hat B}^\theta_L M \hat{B}^\theta_R \Sigma^\dagger] + \bar{a}_1 \Tr[\bar{\hat B}^\theta_R \Sigma^\dagger M \Sigma^\dagger \hat B^\theta_L \Sigma] \nn
\\
&+ a_2 \Tr[\bar{\hat B}^\theta_R \Sigma^\dagger \hat B^\theta_L M] + \bar{a}_2 \Tr[\bar{\hat B}^\theta_L \Sigma \hat B^\theta_R \Sigma^\dagger M \Sigma^\dagger] \nn
\\
&+a_3 \Tr[\bar{\hat B}^\theta_L \Sigma \hat B^\theta_R \Sigma^\dagger + \bar{\hat B}^\theta_R \Sigma^\dagger \hat B^\theta_L \Sigma] \Tr[\Sigma^\dagger M] +\hc \,, \label{linearmassbaryon}
\\
\mathcal{L}^0_{ \ell} &=  \sum_{\ell = e,\mu}\bar{\ell}( i\gamma^\mu \partial_\mu-m_\ell) \ell\,.
\end{align}
where we recall the quark mass matrix spurion transforms as $M \to L M R^\dagger$, we dropped some terms at the same order in derivatives (acting on the $\Sigma$ matrices) that are irrelevant for our discussion, and we included leptons.

\subsection{Adding chemical potential}

We add chemical potentials for the three mutually commuting abelian symmetries associated with neutron and proton numbers and electric charge (from here on we neglect the other baryons). Under $U(1)_{n,p}$, $\psi \to e^{i \alpha} \psi$ for $\psi = n, p$ respectively, while under $U(1)_{\EM}$ electromagnetism,
\begin{align}
\hat B_{L,R} \to e^{i \alpha Q_e} \hat B_{L,R}  e^{-i \alpha Q_e} \,, \;\; \Sigma \to e^{i \alpha Q_e} \Sigma e^{-i \alpha Q_e} \,, \;\; \ell \to e^{-i \alpha} \ell\,,
\end{align}
with
\begin{align}
Q_e = \frac{1}{3}\begin{pmatrix}
2 && \\
& -1 &\\
&& -1
\end{pmatrix}\,.
\label{hatmuDef1}
\end{align}
Chemical potentials are introducing following the prescription in \Eq{chemicalPotental}, i.e.~by modifying temporal derivatives as
\begin{align}
\partial_0 \Sigma & \to \partial_0 \Sigma + i [\hat{\mu},\Sigma]\,,
\\
\partial_0 \hat B_{L,R} &\to \partial_0 
\hat B_{L,R} + i [\hat{\mu},\hat B_{L,R}] + i \hat{\mu}_{n,p} \hat B_{L,R} \,,
\\
\partial_0 \ell &\to \partial_0 \ell - i \mu \ell\,,
\end{align}
where we denoted
\begin{align}
\hat{\mu} = \mu \, Q_e\,, \;\;\; \hat{\mu}_{n,p} = \text{Diag}[\mu_p-\mu,\mu_n,0] \,. 
\label{hatmuDef2}
\end{align}
We then get the following additional terms to the Lagrangian (\ref{linearL})
\begin{align}
\mathcal{L}^\mu_{\Sigma}&= \mathcal{L}^0_{\Sigma} + \frac{f_\pi^2}{4} \left( \Tr[2 i \partial_0 \Sigma [\hat{\mu},\Sigma^\dagger]] - \Tr[[\hat{\mu},\Sigma][\hat{\mu},\Sigma^\dagger] ] \right) \,,
\\
\mathcal{L}^\mu_{B}&= \mathcal{L}^0_{B}- \left( \Tr[\bar{\hat B}^\theta_L \gamma^0 [\hat{\mu}, \hat B^\theta_L]] + \Tr[\bar{\hat B}^\theta_R \gamma^0 [\hat{\mu}, \hat B^\theta_R] ] \right) \nn
\\
&\,\,\,\, \qquad - \left( \Tr [\bar{\hat B}^\theta_L \gamma^0 \hat{\mu}_{n,p} \hat B^\theta_L] + \Tr[ \bar{\hat B}^\theta_R \gamma^0 \hat{\mu}_{n,p} \hat B^\theta_R] \right)\,, \label{swave}
\\
\mathcal{L}^\mu_{ \ell} &= \mathcal{L}^0_{ \ell}+\mu  \sum_{\ell = e,\mu}\bar{\ell} \gamma^0 \ell\,.
\end{align}

\subsection{Non-linear field basis}

It is usually most convenient to work in a field basis for the baryons in which these only transform under the non-linearly realized unbroken $SU(3)$ subgroup of $SU(3)_L \times SU(3)_R$,
\begin{align}
B^\theta_L \to V_\theta B^\theta_L V_\theta^\dagger \,, \;\;\; B^\theta_R = V_\theta B^\theta_R V_\theta^\dagger \,,
\end{align}
with the dressed fields
\begin{align}
B_R^\theta = \xi_R^\dagger \hat B_R^\theta \xi_R \,, \;\;\; B_L^\theta = \xi_L^\dagger \hat B_L^\theta \xi_L \,.
\end{align}
In this basis there are no non-derivative interactions of the mesons with the baryons from the mass terms in \Eq{linearmassbaryon}. 
Besides, in complete analogy to \Eq{bRotate}, the standard (non-rotated) baryons are given by
\begin{align}
&B_L = \xi_0^\dagger B_L^\theta \xi_0 \,, \;\;\; B_R= \xi_0 B_R^\theta \xi_0^\dagger \,.
\end{align}
In terms of such fields, which we recall make up the finite-density background, the baryon Lagrangian is given by
\begin{align}
\mathcal{L}_B^\mu &= i \Tr [\bar{B} \gamma^\mu D_\mu B] - M_B \Tr[ \bar{B} B] - \mu \Tr[ \bar{B} \gamma^0 [\hat Q_e,B]] - \Tr [ \bar{B} \gamma^0 \hat{\mu}_{u,d} B]\,,
\end{align}
where the baryon covariant derivative is given by $D_\mu B = \partial_\mu B + [e_\mu,B]$, with 
\begin{align}
e_\mu \equiv \frac12 \left(\xi_0^\dagger (e_L)_{\mu} \xi_0 + \xi_0  (e_R)_{\mu} \xi_0^\dagger\right) \, , \quad 
(e_L)_{\mu} \equiv i \xi_L^\dagger \partial_\mu \xi_L \,, \;\; (e_R)_{\mu} \equiv i \xi_R^\dagger \partial_\mu \xi_R\,,
\end{align}
and 
\begin{align}
\hat Q_e \equiv \frac{1}{2} \left( \xi_0^\dagger  \xi_L^\dagger Q_e \xi_L \xi_0 + \xi_0  \xi_R^\dagger Q_e \xi_R \xi_0^\dagger \right) \, ,
\end{align}
reproducing \Eq{LBdensity}.
The part of the Lagragian proportional to the quark mass matrix reads
\begin{align}
\mathcal{L}_M &= -\frac{\vev{\bar{q}q}_0}{2} \Tr[\hat{M}]  
+ a_1 \Tr[ \bar{B}_L \hat{M} B_R ]+ \bar{a}_1 \Tr[\bar{B}_R \hat{M} B_L  ] \nn
\\
&+ a_2 \Tr[ \bar{B}_R B_L \hat{M}] + \bar{a}_2 \Tr[\bar{B}_L B_R \hat{M} ] \nn
+a_3 \Tr[ \bar{B}_L B_R + \bar{B}_R B_L ] \Tr[\hat{M}] + \hc \,.
\end{align}
where we defined the dressed mass matrix
\begin{align}
\hat{M} &\equiv \xi_0^\dagger \xi_L^\dagger M \xi_R  \xi_0^\dagger\,,
\end{align}
as in \Eq{PotentialNuclear}.
From $\mathcal{L}_M$ in this form it becomes apparent that the $L \leftrightarrow R$ exchange symmetry of QCD implies $a_{1,2} = \bar{a}_{1,2}$, which allows us to write
\begin{align}
\mathcal{L}_M &= -\frac{1}{2} \text{Tr}[\vev{\bar{q} q}_n(\hat{M}+\hat{M}^\dagger)]  \,,
\\
\vev{\bar{q} q}_n &\equiv \vev{\bar{q}q}_0 \mathbb{1}_3 - 2a_1 B\bar{B} - 2a_2 \bar{B}B-2a_3\Tr[\bar{B} B] \mathbb{1}_3 \,.
\end{align}
From this expression one can derive the density-dependent quark condensate of \Eq{qqs}, since in the 
non-relativistic limit $\bar B B = \bar B \gamma_0 B$ and in the mean-field approximation we can treat the baryons as fixed classical background fields, thus
\begin{align}
\bar p p = \bar{p} \gamma_0 p \to \vev{\bar p \gamma_0 p} = n_p
\end{align}
and likewise for the neutron. 
Besides, note that the baryon masses $m_n$ and $m_p$ are given in terms of $\{\sigma_{\pi N},\tilde{\sigma}_{\pi N},\sigma_s\}$ in \Eqs{mnDef}{mpDef} respectively. One can then relate the coefficients $\{a_1,a_2,a_3\}$ of the baryon chiral Lagrangian to the sigma terms
\begin{align}
{\sigma_{\pi N}} &= -2\bar{m}(a_1 +2a_3)\,,
\\
{\tilde{\sigma}_{\pi N}}  &= 2\Delta m\,a_1\,,
\\
{\sigma_s} &= -2m_s(a_2+a_3)\,. \label{sigmaParameters}
\end{align}
Finally, we recall that at zero temperature all the states with $E(p) = \sqrt{p^2+m_\psi^2}< \mu_\psi$ are occupied, such that
\begin{align}
n_\psi =  \vev{\bar{\psi}(x) \gamma_0 \psi(x)} = g_\psi \int_{0}^{E(p)< \mu_\psi} \frac{\mathrm{d}^3p}{(2\pi)^3}  = \frac{g_\psi }{6\pi^2} (\mu_\psi^2-m_\psi^2)^{3/2}\,,
\end{align}
with $g_\psi$ counting the internal degrees of freedom, e.g.~$ g_\psi=2$ for a fermion. In \Sec{qqFiniteDensitySubSection} we fixed the values of $\{n ,n_p\}$ by implicitly fixing the values of $\{\mu_p,\mu_n\}$
\begin{align}
\mu_p  &=\sqrt{(3\pi^2  n_p)^{2/3}+m_p^2}\,,
\\
 \mu_n  &= \sqrt{(3\pi^2 n_n)^{2/3}+m_n^2}\,.
\end{align}
Note that one can fix $\{n ,n_p\}$ while still keeping the charge chemical potential $\mu$ free by choosing the appropriate value of $\mu_p$, namely if $\mu \to \mu+\delta \mu$, then $\mu_p \to \mu_p - \delta\mu$.

\section{Axion mass in Kaon-condensed phase} \label{axionAnalyticalMass}

Tree-level mixing with the mesons in the kaon-condensed phase are removed when the matrix $Q_a$ satisfies the following condition
\begin{align}
\{ \vev{\bar{q} q}_{n}, \xi_0 M Q_a \xi_0+\xi_0^\dagger M Q_a \xi_0^\dagger\} \propto \mathbb{1}_3\,. \label{removeCondition}
\end{align}
If $\text{Re}(\Sigma_0) $ is a diagonal matrix, such that $[\text{Re}(\Sigma_0),\vev{\bar q q}_{n}]=0$, the $Q_a$ matrix given by
\begin{align}
(Q_a)^{\theta}_{n} = \frac{X^\theta_{n}}{\text{Tr}X^\theta_{n}} \, , \;\;\; X^\theta_{n} =  M^{-1 }\left(\xi_0 \frac{\vev{\bar q q}^{-1}_{n}}{\text{Re}(\Sigma_0)} \xi_0^\dagger+\xi_0^\dagger \frac{\vev{\bar q q}^{-1}_{n}}{\text{Re}(\Sigma_0)} \xi_0 \right)\,, \label{QaNoMixing}
\end{align}
satisfies \ref{removeCondition}. Plugging \Eq{QaNoMixing} in \Eq{PotentialNuclear}, we find the axion mass 
\begin{align}
(m_a^2)_{\theta,n} = -\frac{1}{2f_a^2 } \text{Tr}\left[  \vev{\bar{q} q}_{n}\left( \xi_0 M(Q^{\theta}_a)^2 \xi_0+\xi_0^\dagger M(Q^{\theta}_a)^2 \xi_0^\dagger \right)\right]\,.
\end{align}

\bibliography{finiteDensityAxion}
\bibliographystyle{jhep}

\end{document}